\documentclass[journal]{IEEEtran}
\usepackage[dvips]{graphicx}
\usepackage{epsfig}
\usepackage{amsmath}
\usepackage{epsfig}
\usepackage{array}
\usepackage{amssymb}
\usepackage{color}
\newtheorem{Proposition}{\textbf{Proposition}}

\newtheorem{Lemma}{\textbf{Lemma}}
\newtheorem{Theorem}{\textbf{Theorem}}
\newtheorem{Corollary}{\textbf{Corollary}}

\begin{document}

\title{On Secret-Message Transmission by Echoing Encrypted Probes}

\author{Yingbo Hua, \emph{Fellow, IEEE}
\thanks{Department of Electrical and Computer Engineering,
University of California, Riverside, CA 92521, USA. Email: yhua@ece.ucr.edu. This work was supported in part by the Department of Defense under W911NF-20-2-0267. The views and conclusions contained in this
document are those of the author and should not be interpreted as representing the official policies, either
expressed or implied, of  the U.S. Government. The U.S. Government is
authorized to reproduce and distribute reprints for Government purposes notwithstanding any copyright
notation herein.}
}

\maketitle

\begin{abstract}
A scheme for secure communications, called ``Secret-message Transmission by Echoing Encrypted Probes (STEEP)'', is revisited. STEEP is a round-trip scheme with a probing phase from one user to another and an echoing phase in the reverse direction. STEEP is shown to be broadly applicable to yield a positive secrecy rate in bits per channel use even if the receive channels at eavesdropper (Eve) are stronger than those between legitimate users in both forward and reverse directions.   This paper focuses on STEEP in the following settings: using Gaussian probing signal and Gaussian linear encryption over MIMO Gaussian channel (G-STEEP); using phase-shift-keying probing signal and a nonlinear encryption over SISO channel (P-STEEP); and a variation of G-STEEP for multiple access communication (M-STEEP). In each of the settings, Eve is assumed to have any given number of antennas, and STEEP is shown to yield a positive secrecy rate subject to a sufficiently large power in the echoing phase, as long as Eve's receive channel in the probing phase is not noiseless. It is also shown that G-STEEP, subject to asymmetric large powers in forward and reverse directions, has its secrecy rate approaching the secret-key capacity based on Gaussian probing signal over MIMO Gaussian channel. STEEP does not require secure feedback channel, collaborative third party, in-band full-duplex or reciprocal channels between users, but only needs a design for echoing encrypted probes, asymmetric power allocation and/or collaborative round-trip coding.
\end{abstract}


\section{Introduction}

Secret-message transmission from one node to another subject to eavesdropping has been a long-standing problem for secure communications, which is encountered widely in modern networks. The information-theoretical study of this problem, nowadays known as physical layer security, has a long history since Shannon's work \cite{Shannon1949} in 1940's. Comprehensive reviews of this subject are available in \cite{Bloch2011}, \cite{Poor2017} and \cite{Zhang2020} among others. Many achievements of great importance have been made by researchers in this field, which are centered around wiretap channel (WTC) and secret key generation (SKG). Yet, to the author's knowledge, few among the numerous works on WTC developed since \cite{Wyner1975} and \cite{Csiszar1978} in 1970's could tell us how to produce a positive secrecy rate between Alice and Bob when the channel between them is half-duplex and always weaker than the receive channel at an eavesdropper (Eve). And few among the numerous works on SKG developed since \cite{Maurer1993} and \cite{Ahlswede1993} in 1990's could tell us how to connect  their developments to a WTC scheme in a broadly beneficial way. There appears a non-negligible disconnect between the numerous works on WTC and those on SKG.

A notable exception is however the work in \cite{Hayashi2020} where a two-way protocol using binary signalling over a Gaussian channel is proposed to achieve a positive secrecy rate even if Eve's channel is stronger than users'.
%This protocol is similar to one previously shown in  \cite{Maurer1993} and  revisited in \cite{Gamal2011}.
In fact, there is
a general principle that predates and underpins this protocol, which consists of two integral steps:

First, if Alice transmits independent realizations of a random integer $X$ over a (memoryless) WTC system, and Bob and Eve receive the corresponding realizations of the random integers (binary or not) $Y$ and $Z$, then it is known  \cite{Bloch2011} that the secret-key capacity $C_{key}$ in bits per realization of $\{X,Y,Z\}$ achievable by Alice and Bob via public communications satisfies
\begin{equation}\label{eq:key}
  I(X;Y)-I(Y;Z)\leq C_{key}\leq I(X;Y|Z),
\end{equation}
where $I(X;Y|Z)$ (for example) denotes the mutual information between $X$ and $Y$ conditional on $Z$. The left and right sides of \eqref{eq:key} are known as Maurer's lower and upper bounds \cite{Maurer1993}. In some cases (such as when $Y$ and $Z$ are independent of each other conditioned on $X$), the upper and lower bounds coincide, i.e., $C_{key}=I(X;Y)-I(Y;Z)= I(X;Y|Z)$, which is generally positive regardless of the WTC secrecy rate $[I(X;Y)-I(X;Z)]^+$ from Alice to Bob. Here $x^+\doteq\max(x,0)$. Note that $C_{key}$ is commonly referred to as a ``capacity'' despite the fact that it depends on the distributions of $X$, $Y$ and $Z$ that could be controllable in some applications.

Second, given the random integers $X$, $Y$ and $Z$ at Alice, Bob and Eve respectively, an encryption lemma (see section 4.2.1 in \cite{Bloch2011} or section 22.4.3 in \cite{Gamal2011}) says that Bob can choose a uniform random integer $S$ and transmit $S\oplus Y$ (a modulo sum of $S$ and $Y$) via a public channel so that the secrecy rate of the effective WTC system from Bob to Alice equals $I(X;Y)-I(Y;Z)$.

The above two-step principle is also a foundation for a scheme called  ``Secret-message Transmission by Echoing Encrypted Probes (STEEP)'' \cite{Hua2023Sept}. However, differing from \cite{Maurer1993}, \cite{Bloch2011}, \cite{Gamal2011} and \cite{Hayashi2020}, STEEP as shown in this paper allows the following extensions: $X$, $Y$ and $Z$ are allowed to be any in the spaces of real and/or complex numbers and vectors and/or matrices; the modulo sum $\oplus$ is allowed to be replaced by other suitable operations (examples will be shown); and the public channel from Bob to Alice and Eve may be replaced by any channels at the physical (or an upper) layer. Not necessarily all optimal in information theory, these extensions allow secure communications in a wider range of settings to be conducted rather simply with a guaranteed positive secrecy rate.

\begin{figure}[ht]
  \centering
  \includegraphics[width=2.3in]{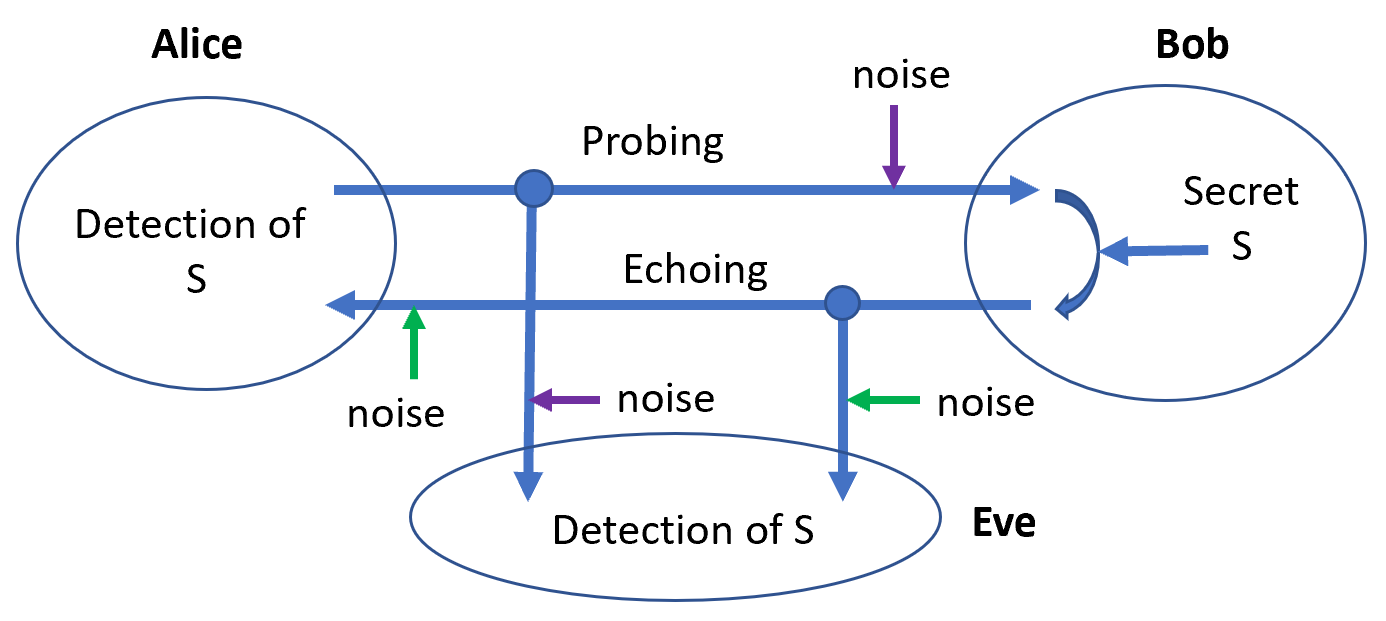}
  \caption{STEEP is a round-trip scheme with embedded secret message on returned (estimated) probes. There is no requirement of secure feedback or in-band full duplex. The same probing signal transmitted/broadcasted by AP in phase 1 could be used by multiple users for orthogonal multiple access in phase 2.}\label{Fig:STEEP}
\end{figure}

As illustrated in Fig. \ref{Fig:STEEP}, there are two collaborative phases in STEEP. In phase 1 (or probing phase), random symbols or probes are transmitted from Alice to Bob. These probes arrive at Bob after a transformation by the channel response, which could result in some ``effective'' probes that can be estimated consistently (but not necessarily perfectly) by Bob. The exact definition of ``effective probe'' may vary, depending on how STEEP is implemented.
In phase 2 (or echoing phase) of STEEP,  Bob's estimates of the effective probes are encrypted or combined with secret message symbols before they are transmitted (``echoed'' back) to Alice. These collaborative two-phase operations result in an effective WTC system  from Bob to Alice and Eve, which is almost surely in favor of the users subject to a sufficient power from Bob.

Evolved from a scheme (called iSAT) shown in \cite{Hua2023}, STEEP is a collaborative round-trip scheme between half-duplex nodes, which has a broad applicability and differs from many two-way full-duplex schemes in the literature such as \cite{Lai2008}, \cite{Khisti2012}, \cite{HeYener2013} and \cite{HuaLiang2023}. A latest work in \cite{Hayashi2023} also assumes two-way full-duplex to explore the fundamental limits on the secrecy rate region between two users with little consideration of complexity and practicality. The scheme in \cite{Hayashi2020}  however can be seen as a special case of STEEP subject to binary symmetric channels.

The goal of this paper is to present STEEP in its latest forms.
The primary contributions include novel insights into STEEP in three different settings. The first setting (or G-STEEP) uses Gaussian probing signal (GPS) and Gaussian linear encryption (GLE) over MIMO channels between two users, for which an achievable secrecy rate $R_{s,G}$  is derived and analyzed. In particular, $R_{s,G}$ is shown to converge to the secret-key capacity $C_{key}$ based on GPS over MIMO channels if the echoing power in G-STEEP dominates the probing power and both become large.
The second setting (or P-STEEP) uses phase-shift-key (PSK) probing signal and PSK nonlinear encryption between two users, for which an achievable secrecy rate $R_{s,P}$ is also presented. The third setting (or M-STEEP) uses GPS and GLE over multiple access channels between an access point (AP) and multiple users all of whom exploit the same probes from the AP. An achievable secrecy rate $\tilde R_{s,1}$ of M-STEEP from an arbitrary user to AP is shown to be a function that decreases gradually with some robustness (instead of abruptly) as the number $M$ of users increases. In each setting, the achievable secrecy rate of STEEP is shown to be positive subject to a sufficiently large power in the echoing phase, which includes the secrecy rate from a user to AP subject to exposure of messages from all other users.

The paper is organized as follows. The physical-layer channel models of interest in this paper are described in section \ref{sec:channel_model}, which also highlights some prior results and the main problem of interest in this paper.  G-STEEP, P-STEEP and M-STEEP are presented and analyzed respectively in sections \ref{sec:G-STEEP}, \ref{sec:P-STEEP} and \ref{sec:M-STEEP}.  Much of the technical proofs is relegated to the appendix. The paper is ended with additional comments and conclusion.

%Notations: Matrices are denoted by bold-face capital letters, vectors by bold-face lower case letters, determinant and F-norm of $\mathbf{A}$ by $|\mathbf{A}|$ and $\|\mathbf{A}\|$, expectation by $\mathbb{E}$, trace by $\texttt{Tr}$, circular complex Gaussian distribution with mean vector $\mathbf{m}$ and covariance matrix $\mathbf{C}$ by $\mathcal{CN}(\mathbf{m},\mathbf{C})$, the set of all $n\times m$ complex matrices by $\mathbb{C}^{n\times m}$. 

\section{Channel Models, Prior Results, and The Problem}\label{sec:channel_model}
\subsection{Channel model}
We will first consider a three-node network with two legitimate users (Alice and Bob) and an eavesdropper (Eve). The numbers of antennas on them are denoted respectively by $n_A$, $n_B$ and $n_E$.  In the case of wireline communications, each antenna here corresponds to a transceiver.

When Alice transmits (within a coherence time $\mathcal{T}_1$) a sequence of random vectors $\sqrt{\frac{p_A}{n_A}}\mathbf{x}_A(k)\in\mathbb{C}^{n_A\times 1}$ of power $p_A$, we assume that Bob and Eve receive respectively
\begin{align}\label{eq:yBkp}
  &\mathbf{y}_B(k)=\sqrt{p_A/n_A}\mathbf{H}_{BA}\mathbf{x}_A(k)+\mathbf{w}_B(k),
\\\label{eq:yEAkp}
  &\mathbf{y}_{EA}(k)=\sqrt{p_A/n_A}\mathbf{H}_{EA}\mathbf{x}_A(k)+\mathbf{w}_{EA}(k).
\end{align}
where all noise entries are mutually independent with the normalized Gaussian distribution, i.e., $\mathbf{w}_B(k)$ is $\mathcal{CN}(0,\mathbf{I}_{n_B})$ and $\mathbf{w}_{EA}(k)$ is $\mathcal{CN}(0,\mathbf{I}_{n_E})$. For notational simplicity, we will also use the scaled versions of $\mathbf{H}_{BA}$ and $\mathbf{H}_{EA}$, i.e., $\mathbf{H}_{BA}'\doteq\sqrt{p_A/n_A}\mathbf{H}_{BA}\in\mathbb{C}^{n_B\times n_A}$ and $\mathbf{H}_{EA}'\doteq\sqrt{p_A/n_A}\mathbf{H}_{EA}\in\mathbb{C}^{n_E\times n_A}$. Here $k$ is the sampling index.

Similarly,
when Bob transmits (within a coherence time $\mathcal{T}_2$) a sequence of random vectors $\sqrt{\frac{p_B}{n_B}}\mathbf{x}_B(k)\in\mathbb{C}^{n_B\times 1}$ of power $p_B$, we assume that Alice and Eve receive respectively
\begin{align}\label{eq:yAk}
  &\mathbf{y}_A(k)=\sqrt{p_B/n_B}\mathbf{H}_{AB}\mathbf{x}_B(k)+\mathbf{w}_A(k),
\\\label{eq:yEBk}
  &\mathbf{y}_{EB}(k)=\sqrt{p_B/n_B}\mathbf{H}_{EB}\mathbf{x}_B(k)+\mathbf{w}_{EB}(k),
\end{align}
where the normalized noises $\mathbf{w}_A(k)$ and $\mathbf{w}_{EB}(k)$ are  $\mathcal{CN}(\mathbf{0},\mathbf{I}_{n_A})$ and $\mathcal{CN}(\mathbf{0},\mathbf{I}_{n_E})$. We will also write
$\mathbf{H}_{AB}'=\sqrt{p_B/n_B}\mathbf{H}_{AB}\in\mathbb{C}^{n_A\times n_B}$ and $\mathbf{H}_{EB}'=\sqrt{p_B/n_B}\mathbf{H}_{EB}\in\mathbb{C}^{n_E\times n_B}$.

Alice and Bob are half-duplex (unless indicated otherwise). Namely, $\mathcal{T}_1$ and $\mathcal{T}_2$ do not overlap. But $\mathcal{T}_1$ and $\mathcal{T}_2$ may or may not belong to a common coherence period.

Every receive channel parameter is assumed to be known to the corresponding receiver. If there is any required feedback of channel parameters between users, these parameters are also assumed to be known to Eve. In fact, all channel parameters in this paper are treated as known to Eve.

Also assume that all signals and noises in each transmission direction (i.e., from Alice to Bob, or from Bob to Alice) are temporally independent.
So, for simpler notations, we will also drop the sampling (or slot) index ``$k$''.
In this case, one should view the channel matrices as constant but the transmitted signals (and the noises) as random. The results on secrecy rates will be based on a large number of slots in each of probing and echoing phases.
In the case of temporally coded transmissions, the assumption of ``temporal independence'' could typically serve as an approximation.

In section \ref{sec:M-STEEP}, we will also consider an orthogonal multiple access problem where the access point (AP) has $n_A$ antennas and each of $M$ user equipment (UEs) has a single antenna. The channel parameters and noises are similarly defined.

\subsection{Some prior results and the problem}
In the classic WTC scheme, the signals transmitted from Alice to Bob and Eve are not coordinated with any signals transmitted from Bob to Alice and Eve (even though both ends of a physical link are typically able to transmit).  In this case, assuming all channel parameters are public, the secrecy capacity from Alice to Bob (in bits per complex channel use) is known \cite{KhistiWornell2010,OggierHassibi2011} to be
\begin{equation}\label{eq:classic_WTC}
  C_{s,A\to B}=\max_{\mathbf{K}_x,\texttt{Tr}\{\mathbf{K}_x\}\leq n_A} \log\frac{|\mathbf{I}_{n_B}+\frac{p_A}{n_A}\mathbf{H}_{B,A}\mathbf{K}_x\mathbf{H}_{B,A}^H|}
  {|\mathbf{I}_{n_E}+\frac{p_A}{n_A}\mathbf{H}_{E,A}\mathbf{K}_x\mathbf{H}_{E,A}^H|}
\end{equation}
where $\mathbf{K}_x=\mathbb{E}\{\mathbf{x}_A(k)\mathbf{x}_A^H(k)\}$. (The result for $C_{s,B\to A}$ would be obvious.) This secrecy capacity is achieved by a Gaussian codebook, i.e., $\mathbf{x}_A(k)$ follows the circular complex Gaussian distribution $\mathcal{CN}(\mathbf{0},\mathbf{K}_x)$. Furthermore, $C_{s,A\to B}>0$ \cite{KhistiWornell2010} if and only if (regardless of the positive power $p_A$ and, of course, $p_B$)
\begin{equation}\label{eq:channel}
 \alpha\doteq \min_{\mathbf{v}\in\mathcal{C}^{n_A\times 1}}\frac{\|\mathbf{H}_{E,A}\mathbf{v}\|^2}{\|\mathbf{H}_{B,A}\mathbf{v}\|^2}<1.
\end{equation}
This means that Eve's receive channel from Alice must be weaker than Bob's receive channel from Alice in order for the classic WTC scheme to yield a positive secrecy rate from Alice to Bob. The above condition is however not always feasible. Specifically, when $n_E\geq n_A$, $\alpha$ is very likely larger than one in many practical situations especially where Eve is closer to Alice than Bob is.

Note that if $C_{s,A\to B}>0$, it is achieved by $\texttt{Tr}\{\mathbf{K}_x\}= n_A$ (using full power) \cite{OggierHassibi2011}. But  if $C_{s,A\to B}=0$, then it is obviously achieved by $\texttt{Tr}\{\mathbf{K}_x\}=0$ (using zero power), but not necessarily by $\texttt{Tr}\{\mathbf{K}_x\}= n_A$.

The main problem of interest in this paper is how to achieve a positive secrecy rate between two users (Alice and Bob), and between an access point and multiple user equipment, even if Eve's receive channel is stronger than users'. In particular, we aim to present novel insights into STEEP and to reveal its ability to achieve a positive secrecy rate under virtually all channel conditions.

Subject to Gaussian distributed $\mathbf{x}_A(k)$ and any given $\mathbf{K}_x$, it is already established that there is a WTC coding scheme to yield a secrecy rate (see \cite{Bloch2011} among many sources):
\begin{align}\label{eq:classic_WTC_2}
  &R_{s,A\to B}=\left (I(\mathbf{x}_A(k);\mathbf{y}_B(k))- I(\mathbf{x}_A(k);\mathbf{y}_{EA}(k))\right )^+\notag\\
  &= \left (\log\frac{|\mathbf{I}_{n_B}+\frac{p_A}{n_A}\mathbf{H}_{B,A}\mathbf{K}_x\mathbf{H}_{B,A}^H|}
  {|\mathbf{I}_{n_E}+\frac{p_A}{n_A}\mathbf{H}_{E,A}\mathbf{K}_x\mathbf{H}_{E,A}^H|}\right )^+\leq C_{s,A\to B},
\end{align}
where the equality holds when $n_A=1$.
This result or its equivalent form will be used later for a number of channel conditions including virtual channel conditions. In particular, STEEP is a strategy that transforms a physical channel condition, even when \eqref{eq:channel} is not satisfied, into a virtual channel condition for which a positive secrecy rate can be ensured by a power control (i.e., collaboratively controlling $p_A$ and $p_B$).

\section{STEEP with Gaussian Channel Probing and Gaussian Linear Encryption (G-STEEP)}\label{sec:G-STEEP}
In this section, G-STEEP is presented, and an achievable secrecy rate $R_{s,G}$ of G-STEEP is then derived and discussed. Properties of $R_{s,G}$ subject to large powers are highlighted.

\subsection{Description of G-STEEP}
In phase 1 of G-STEEP, Alice applies Gaussian probing signal, i.e., she transmits a realization of the random probing vector $\sqrt{\frac{p_A}{n_A}}\mathbf{x}_A$ in each probing slot where $\mathbf{x}_A$ is assumed to be $\mathcal{CN}(0,\mathbf{I}_{n_A})$.
 The corresponding signal received by Bob is $\mathbf{y}_B$ in \eqref{eq:yAk}, i.e., after dropping ``$k$'',
 \begin{equation}\label{}
   \mathbf{y}_B=\mathbf{H}_{BA}'\mathbf{x}_A+\mathbf{w}_B.
 \end{equation}

 Note that given $n_A\geq n_B$, the probing phase should be from Alice to Bob in order to have the largest $R_{s,G}$. This is because the degree of freedom (DoF) of the secret-key capacity $C_{key}$ based on channel probing from a node with more antennas is larger than that  from a node with less antennas. See \cite{Hua2023} and \cite{HuaMaksud2024}. Such a connection between $R_{s,G}$ and $C_{key}$ will also be shown.

 In phase 2 of G-STEEP, Bob transmits his estimated probing vector subject to a Gaussian linear encryption, i.e., he transmits $\sqrt{\frac{p_B}{2n_B}}\mathbf{x}_B=\sqrt{\frac{p_B}{2n_B}}(\mathbf{\hat p}+\mathbf{s})$ where $\mathbf{\hat p}$ is his estimate of the probing vector and $\mathbf{s}$ is a secret-message dependent vector and assumed to be $\mathcal{CN}(0,\mathbf{I}_{n_B})$. Here $p_B$ is the upper bound of the total transmit power from Bob.  The corresponding signal received by Alice is
\begin{equation}\label{eq:yA}
  \mathbf{y}_A=\mathbf{H}_{AB}''(\mathbf{\hat p}+\mathbf{s})+\mathbf{w}_A
\end{equation}
with $\mathbf{H}_{AB}''=\sqrt{1/2}\mathbf{H}_{AB}'=\sqrt{p_B/(2n_B)}\mathbf{H}_{AB}$.

The above protocol creates an effective wiretap channel (eWTC) system from $\mathbf{s}$ at Bob and the optimal estimate of $\mathbf{s}$ at Alice and the optimal estimate of $\mathbf{s}$ at Eve. We will show that this eWTC is in general in favor of the users.

\subsection{Analysis of the signal received by Bob in Phase 1}

Let the (thin) SVD of $\mathbf{H}_{BA}$ be $\mathbf{U}_{BA}\boldsymbol{\Pi}_{BA}\mathbf{V}_{BA}^H
=\sum_{i=1}^{n_B}\pi_{BA,i}\mathbf{u}_{BA,i}\mathbf{v}_{BA,i}^H$ where $\mathbf{U}_{BA}=[\mathbf{u}_{BA,1},\cdots,\mathbf{u}_{BA,n_B}]\in\mathbb{C}^{n_B\times n_B}$ is unitary and  $\mathbf{V}_{BA}=[\mathbf{v}_{BA,1},\cdots,\mathbf{v}_{BA,n_B}]\in\mathbb{C}^{n_A\times n_B}$ is column-wise orthonormal. Then we can write
\begin{align}\label{}
  &\mathbf{y}_B =\mathbf{U}_{BA}\boldsymbol{\Pi}_{BA}'\mathbf{p} +\mathbf{w}_B
\end{align}
with $\boldsymbol{\Pi}_{BA}'=\sqrt{p_A/n_A}\boldsymbol{\Pi}_{BA}$.
Here $\mathbf{p}\doteq\mathbf{V}_{BA}^H\mathbf{x}_A$ which is here by definition the \emph{effective probing vector} at Bob. Clearly, $\mathbf{p}$ is $\mathcal{CN}(0,\mathbf{I}_{n_B})$ given $\mathbf{x}_A$ being $\mathcal{CN}(0,\mathbf{I}_{n_A})$.

Assume that Alice and Eve both know the feedback of $\mathbf{V}_{BA}$ from Bob. Then, Alice also knows the effective probing vector $\mathbf{p}=\mathbf{V}_{BA}^H\mathbf{x}_A$. But if $n_A=n_B=1$, there is no need for feedback of $\mathbf{V}_{BA}$.

Given the Gaussian signal and noise model, the MMSE (minimum-mean-squared-error) estimate $\mathbf{\hat p}$ of $\mathbf{p}$ by Bob is linear and given by
\begin{align}\label{}
  &\mathbf{\hat p}=\mathbb{E}\{\mathbf{p}\mathbf{y}_B^H\}(\mathbb{E}\{\mathbf{\mathbf{y}_B}\mathbf{y}_B^H\})^{-1}
  \mathbf{y}_B\notag\\
  &=
  \boldsymbol{\Pi}_{BA}'\mathbf{U}_{BA}^H
  \left (\mathbf{U}_{BA}\boldsymbol{\Pi}_{BA}'^2\mathbf{U}_{BA}^H+\mathbf{I}_{n_B} \right )^{-1}\mathbf{y}_B\notag\\
  &=\boldsymbol{\Pi}_{BA}'
  \left (\boldsymbol{\Pi}_{BA}'^2+\mathbf{I}_{n_B} \right )^{-1}\mathbf{U}_{BA}^H\mathbf{y}_B,
\end{align}
and $\mathbf{R}_{\mathbf{\hat p}}\doteq\mathbb{E}\{\mathbf{\hat p}\mathbf{\hat p}^H\}=\boldsymbol{\Pi}_{BA}'^2(\boldsymbol{\Pi}_{BA}'^2
+\mathbf{I}_{n_B})^{-1}$. The operator $\mathbb{E}$ denotes the expectation.
The MSE matrix of $\mathbf{\hat p}$ is
\begin{align}\label{eq:R_Delta_p}
  &\mathbf{R}_{\Delta \mathbf{p}} \doteq\mathbb{E}\{(\mathbf{\hat p}-\mathbf{p})(\mathbf{\hat p}-\mathbf{p})^H\}=-\mathbb{E}\{(\mathbf{\hat p}-\mathbf{p})\mathbf{p}^H\}\notag\\
  %&= \mathbf{I}_{n_B}
%  -\frac{p_A}{n_A}\boldsymbol{\Pi}_{BA}\mathbf{U}_{BA}^H\notag\\
%  &\,\,\cdot
%  \left (\frac{p_A}{n_A}\mathbf{U}_{BA}\boldsymbol{\Pi}_{BA}^2\mathbf{U}_{BA}^H+
  %\sigma_B^2\mathbf{I}_{n_B} \right )^{-1}\mathbf{U}_{BA}\boldsymbol{\Pi}_{BA}\notag\\
  &=\mathbf{I}_{n_B}
  -\boldsymbol{\Pi}_{BA}'
  \left (\boldsymbol{\Pi}_{BA}'^2+\mathbf{I}_{n_B} \right )^{-1}\boldsymbol{\Pi}_{BA}'\notag\\
  &=\mathbf{I}_{n_B}
  -\boldsymbol{\Pi}_{BA}'^2
  \left (\boldsymbol{\Pi}_{BA}'^2+\mathbf{I}_{n_B} \right )^{-1}\notag\\
  &=\left (\boldsymbol{\Pi}_{BA}'^2+\mathbf{I}_{n_B} \right )^{-1}=\mathbf{I}_{n_B}-\mathbf{R}_{\mathbf{\hat p}}
\end{align}
which is diagonal with the $i$th diagonal element being $\frac{1}{\pi_{BAi}'^2+1}=\mathcal{O}(1/p_A)$.

\subsection{Analysis of the signal received by Alice in phase 2}

The MMSE estimate of $\mathbf{s}$ by Alice from $\mathbf{y}_A$ (and from her knowledge of the exact $\mathbf{x}_A$) can be based on this zero-mean sufficient statistic $\Delta\mathbf{y}_A\doteq\mathbf{y}_A-\mathbb{E}\{\mathbf{y}_A|\mathbf{x}_A\}$, which can be shown to be
\begin{equation}\label{}
  \Delta\mathbf{y}_A=\mathbf{y}_A-\mathbf{H}_{AB}''\mathbf{R}_{\mathbf{\hat p}}\mathbf{p}.
\end{equation}

Then the MMSE estimate of $\mathbf{s}$ by Alice is
\begin{align}\label{eq:s_hat_A}
  &\mathbf{\hat s}_A = \mathbf{H}_{AB}''^H
  \left (\mathbf{H}_{AB}''(\mathbf{R}_{ \Delta \mathbf{ p}'}+\mathbf{I}_{n_B})\mathbf{H}_{AB}''^H+
  \mathbf{I}_{n_A}
  \right )^{-1}\Delta\mathbf{y}_A.
\end{align}
Here
\begin{equation}\label{}
  \Delta\mathbf{p}'\doteq\mathbf{\hat p}-\mathbf{R}_{\mathbf{\hat p}}\mathbf{p}
=\mathbf{R}_{\Delta \mathbf{p}}\mathbf{\hat p}+\mathbf{R}_{\mathbf{\hat p}}\Delta\mathbf{p},
\end{equation}
% $\Delta\mathbf{p}'\doteq\mathbf{\hat p}-\mathbf{R}_{\mathbf{\hat p}}\mathbf{p}
%=\mathbf{R}_{\Delta \mathbf{p}}\mathbf{\hat p}+\mathbf{R}_{\mathbf{\hat p}}\Delta\mathbf{p}$.
and
\begin{align}\label{eq:RDpP}
  &\mathbf{R}_{\Delta\mathbf{p}'}\doteq\mathbb{E}\{\Delta\mathbf{p}'\Delta\mathbf{p}'^H\}\notag\\
  &=
\mathbf{R}_{\Delta\mathbf{ p}}\mathbf{R}_{ \mathbf{\hat p}}\mathbf{R}_{\Delta\mathbf{p}}+\mathbf{R}_{\mathbf{\hat p}}\mathbf{R}_{\Delta\mathbf{p}}\mathbf{R}_{\mathbf{\hat p}}=\mathbf{R}_{\mathbf{\hat p}}\mathbf{R}_{\Delta\mathbf{p}}.
\end{align}

Then
the MSE matrix of $\mathbf{\hat s}_A$ is
 \begin{align}\label{eq:RDsA}
   &\mathbf{R}_{\Delta\mathbf{s}_A}\doteq\mathbb{E}\{(\mathbf{\hat s}_A-\mathbf{s}_A)(\mathbf{\hat s}_A-\mathbf{s}_A)^H\} \notag\\
   &= \mathbf{I}_{n_B}
   -\mathbf{H}_{AB}''^H\left (\mathbf{H}_{AB}''(\mathbf{R}_{\Delta \mathbf{p}'}+\mathbf{I}_{n_B})\mathbf{H}_{AB}''^H+
  \mathbf{I}_{n_A}
  \right )^{-1}\mathbf{H}_{AB}''\notag\\
  &=\mathbf{I}_{n_B}
   -\left (\mathbf{H}_{AB}''^H\mathbf{H}_{AB}''(\mathbf{R}_{ \Delta \mathbf{p}'}+\mathbf{I}_{n_B})+
  \mathbf{I}_{n_B}
  \right )^{-1}\mathbf{H}_{AB}''^H\mathbf{H}_{AB}''\notag\\
  &=\left (\mathbf{H}_{AB}''^H\mathbf{H}_{AB}''(\mathbf{R}_{ \Delta\mathbf{ p}'}+\mathbf{I}_{n_B})+
  \mathbf{I}_{n_B}
  \right )^{-1}\notag\\
  &\,\,\cdot\left (\mathbf{H}_{AB}''^H\mathbf{H}_{AB}''\mathbf{R}_{ \Delta\mathbf{ p}'}+
  \mathbf{I}_{n_B}
  \right ).
 \end{align}

\subsection{Effective channel capacity from Bob to Alice}
\begin{Lemma}
If $\mathbf{x}\in\mathbb{C}^{n_x}$ and $\mathbf{y}\in\mathbb{C}^{n_y}$ are joint (non-singular) circular complex Gaussian with zero means and covariance matrices $\mathbf{R}_x$ and $\mathbf{R}_y$ respectively, then $I(\mathbf{x};\mathbf{y})=I(\mathbf{x};\mathbf{\hat x})=\log\frac{|\mathbf{R}_x|}{|\mathbf{R}_{x|y}|}$, where $\mathbf{\hat x}$ is the MMSE estimate of $\mathbf{x}$ from $\mathbf{y}$, and $\mathbf{R}_{x|y}$ is the MSE matrix of $\mathbf{\hat x}$.
\end{Lemma}
\begin{IEEEproof}
Let $\mathbf{z}=[\mathbf{x}^T,\mathbf{y}^T]^T$. Then $\mathbf{z}$ is Gaussian with the covariance matrix
\begin{equation}\label{}
  \mathbf{R}_z = \left [\begin{array}{cc}
                          \mathbf{R}_x & \mathbf{R}_{xy} \\
                          \mathbf{R}_{xy}^H & \mathbf{R}_y
                        \end{array}
   \right ].
\end{equation}
It follows that the PDF of $\mathbf{x}$ given $\mathbf{y}$ is
\begin{align}\label{}
  &f(\mathbf{x}|\mathbf{y})=\frac{f(\mathbf{x},\mathbf{y})}{f(\mathbf{y})}
  =\frac{\frac{1}{\pi^{n_x+n_y}|\mathbf{R}_z|}\exp(-\mathbf{z}^H\mathbf{R}_z^{-1}\mathbf{z})}
  {\frac{1}{\pi^{n_y}|\mathbf{R}_y|}\exp(-\mathbf{y}^H\mathbf{R}_y^{-1}\mathbf{y})}
\end{align}
Using the block matrix properties of $\mathbf{R}_z^{-1}$ and $|\mathbf{R}_z|$, one can verify that
\begin{equation}\label{}
  f(\mathbf{x}|\mathbf{y}) = \frac{1}{\pi^{n_x}|\mathbf{R}_{x|y}|}\exp\left (-(\mathbf{x}-\mathbf{\hat x})^H
  \mathbf{R}_{x|y}^{-1}(\mathbf{x}-\mathbf{\hat x})\right )
\end{equation}
with $\mathbf{\hat x} \doteq \mathbf{R}_{xy}\mathbf{R}_y^{-1}\mathbf{y}$ and $\mathbf{R}_{x|y}
\doteq\mathbf{R}_x-
\mathbf{R}_{xy}\mathbf{R}_y^{-1}\mathbf{R}_{xy}^H$. We see that this $\mathbf{\hat x}$ is the MMSE estimate of $\mathbf{x}$ from $\mathbf{y}$ because $\mathbb{E}\{\mathbf{x}|\mathbf{y}\}=\mathbf{\hat x}$, and this $\mathbf{R}_{x|y}$ is the MSE matrix of $\mathbf{\hat x}$.
Finally, we know $I(\mathbf{x};\mathbf{y})=h(\mathbf{x})-h(\mathbf{x}|\mathbf{y})
=\log|\mathbf{R}_x| - \log |\mathbf{R}_{x|y}|$. A constant term in each of the differential entropies $h(\mathbf{x})$ and $h(\mathbf{x}|\mathbf{y})$ canceled each other. We also see that $\mathbf{\hat x}$ is a sufficient statistic of $\mathbf{y}$ for $\mathbf{x}$, and hence $I(\mathbf{x};\mathbf{y})=I(\mathbf{x};\mathbf{\hat x})$.
\end{IEEEproof}

The virtual channel from $\mathbf{s}$ to $\mathbf{\hat s}_A$ is called here the effective channel from Bob to Alice relative to $\mathbf{s}$, the capacity of which (in bits per round-trip symbol interval) is therefore
 \begin{align}\label{eq:CAB_MIMO}
   &C_{A|B,G}\doteq I(\mathbf{s};\{\mathbf{x}_A,\mathbf{y}_A\})= I(\mathbf{s};\mathbf{\hat s}_A)=\log\frac{1}{|\mathbf{R}_{\Delta\mathbf{s}_A}|}\notag\\
  &=\log\frac{|\mathbf{H}_{AB}''^H\mathbf{H}_{AB}''
   (\mathbf{R}_{\Delta\mathbf{p}'}+\mathbf{I}_{n_B})+
  \mathbf{I}_{n_B}|}{|\mathbf{H}_{AB}''^H\mathbf{H}_{AB}''
  \mathbf{R}_{\Delta\mathbf{p}'}+
  \mathbf{I}_{n_B}|}\doteq \log\frac{N_{A|B}}{D_{A|B}}
 \end{align}
 where $N_{A|B}$ and $D_{A|B}$ are defined in the obvious way.
 Here $|A|$ denotes the determinant of $A$. Notice that $\mathbf{s}, \mathbf{x}_A, \mathbf{y}_A$ are jointly Gaussian so that the 2nd and 3rd equalities in \eqref{eq:CAB_MIMO} hold. Note that for $p_A\to 0\texttt{ or }\infty$, $\mathbf{R}_{\Delta\mathbf{p}'}\to \mathbf{0}$ and hence $C_{A|B,G}\to \log|\mathbf{H}_{AB}''^H\mathbf{H}_{AB}''+
  \mathbf{I}_{n_B}|$.

\subsection{Effective channel capacity from Bob to Eve}
To determine the effective channel capacity from Bob to Eve, we need to determine the MSE matrix of the MMSE estimate of $\mathbf{s}$ based on all signals observed by Eve in phases 1 and 2.

After phases 1 and 2 of G-STEEP, the signals received by Eve are
\begin{align}
&\mathbf{y}_{EA}=\mathbf{H}_{EA}'\mathbf{x}_A+\mathbf{w}_{EA},\\
&\mathbf{y}_{EB}=\mathbf{H}_{EB}''(\mathbf{\hat p}
+\mathbf{s})+\mathbf{w}_{EB}.
\end{align}
It follows that
\begin{equation}\label{eq:A2}
  \mathbf{A}\doteq\mathbb{E}\{\mathbf{y}_{EA}\mathbf{y}_{EA}^H\}
  =\mathbf{H}_{EA}'\mathbf{H}_{EA}'^H+\mathbf{I}_{n_E},
\end{equation}
\begin{align}\label{eq:B2}
  &\mathbf{B}\doteq\mathbb{E}\{\mathbf{y}_{EB}\mathbf{y}_{EB}^H\}
  =\mathbf{H}_{EB}''(\mathbf{R}_{\mathbf{\hat p}}+\mathbf{I}_{n_B})\mathbf{H}_{EB}''^H
  +\mathbf{I}_{n_E},
\end{align}
\begin{equation}\label{eq:C2}
  \mathbf{C}\doteq\mathbb{E}\{\mathbf{y}_{EA}\mathbf{y}_{EB}^H\}
  =\mathbf{H}_{EA}'\mathbf{R}_{\mathbf{x}_A,\mathbf{\hat p}}\mathbf{H}_{EB}''^H,
\end{equation}
where $\mathbf{R}_{\mathbf{x}_A,\mathbf{\hat p}}\doteq\mathbb{E}\{\mathbf{x}_A\mathbf{\hat p}^H\}$.
%Here
%\begin{align}\label{eq:R_hat_p}
% & \mathbf{R}_{\mathbf{\hat p}}=\mathbf{I}_{n_B}-\mathbf{R}_{\Delta\mathbf{p}}\notag\\
% &= \frac{p_A}{n_A}\boldsymbol{\Pi}_{BA}
%  \left (\frac{p_A}{n_A}\boldsymbol{\Pi}_{BA}^2+\sigma_B^2\mathbf{I}_{n_B} \right )^{-1}\boldsymbol{\Pi}_{BA}\notag\\
%  &=\frac{p_A}{n_A}\boldsymbol{\Pi}_{BA}^2
%  \left (\frac{p_A}{n_A}\boldsymbol{\Pi}_{BA}^2+\sigma_B^2\mathbf{I}_{n_B} \right )^{-1}.
%\end{align}
Let $\mathbf{Q}\doteq[\mathbf{V}_{BA},\mathbf{V}_{BA}^\perp]\in\mathbb{C}^{n_A\times n_A}$ be a unitary matrix. Then
\begin{align}
&\mathbf{R}_{\mathbf{x}_A,\mathbf{\hat p}}
=\mathbb{E}\{\mathbf{Q}\mathbf{Q}^H\mathbf{x}_A\mathbf{\hat p}^H\}
=\mathbf{Q}\mathbb{E}\left \{\left [\begin{array}{c}
                               \mathbf{V}_{BA}^H\mathbf{x}_A\mathbf{\hat p}^H \\
                               \mathbf{V}_{BA}^{\perp H}\mathbf{x}_A\mathbf{\hat p}^H
                             \end{array}
\right ]\right \}\notag\\
&=\mathbf{Q}\left [ \begin{array}{c}
                      \mathbf{R}_{\mathbf{\hat p}} \\
                      0
                    \end{array}
\right ]=\mathbf{V}_{BA}\mathbf{R}_{\mathbf{\hat p}},
\end{align}
where we have used $\mathbb{E}\{\mathbf{p}\mathbf{\hat p}^H\}=\mathbb{E}\{\mathbf{\hat p}\mathbf{\hat p}^H\}=\mathbf{R}_{\mathbf{\hat p}}$ and $\mathbf{V}_{BA}^{\perp H}\mathbb{E}\{\mathbf{x}_A\mathbf{\hat p}^H\}=0$.

The MMSE estimate of $\mathbf{s}$ by Eve from $\mathbf{y}_{EA}$ and $\mathbf{y}_{EB}$ is
\begin{align}
&\mathbf{\hat s}_E = [\mathbf{0}_{n_B\times n_E},\mathbf{H}_{EB}''^H]\left [ \begin{array}{cc}
                                                                        \mathbf{A} & \mathbf{C} \\
                                                                        \mathbf{C}^H & \mathbf{B}
                                                                      \end{array}
\right ]^{-1}\left [
\begin{array}{c}
  \mathbf{y}_{EA} \\
  \mathbf{y}_{EB}
\end{array}
\right ]
\end{align}
and the MSE matrix of $\mathbf{\hat s}_E$ is
\begin{align}\label{eq:RDsE}
&\mathbf{R}_{\Delta\mathbf{s}_E}\doteq\mathbb{E}\{(\mathbf{\hat s}_E-\mathbf{s})(\mathbf{\hat s}_E-\mathbf{s})^H\}\notag\\
&=\mathbf{I}_{n_B}-
[\mathbf{0}_{n_B\times n_E},\mathbf{H}_{EB}''^H]\left [ \begin{array}{cc}
                                                                        \mathbf{A} & \mathbf{C} \\
                                                                        \mathbf{C}^H & \mathbf{B}
                                                                      \end{array}
\right ]^{-1}\left [\begin{array}{c}
                      \mathbf{0}_{n_E\times n_B} \\
                      \mathbf{H}_{EB}''
                    \end{array}\right ]\notag\\
                    &=\mathbf{I}_{n_B}-\mathbf{H}_{EB}''^H
                    (\mathbf{B}-\mathbf{C}^H\mathbf{A}^{-1}\mathbf{C})^{-1}\mathbf{H}_{EB}''.
\end{align}

It follows from \eqref{eq:B2} and \eqref{eq:C2} that
\begin{align}
&\mathbf{C}^H\mathbf{A}^{-1}\mathbf{C}
=\mathbf{H}_{EB}''\mathbf{T}\mathbf{H}_{EB}''^H
\end{align}
with
\begin{align}\label{eq:T}
  &\mathbf{T}=\mathbf{R}_{\mathbf{x}_A,\mathbf{\hat p}}^H\mathbf{H}_{EA}'^H\left (\mathbf{H}_{EA}'\mathbf{H}_{EA}'^H+\mathbf{I}_{n_E}\right )^{-1}
\mathbf{H}_{EA}'\mathbf{R}_{\mathbf{x}_A,\mathbf{\hat p}}\notag\\
&=\mathbf{R}_{\mathbf{\hat p}}^H\mathbf{V}_{BA}^H\left (\mathbf{H}_{EA}'^H\mathbf{H}_{EA}'+\mathbf{I}_{n_A}\right )^{-1}
\mathbf{H}_{EA}'^H\mathbf{H}_{EA}'\mathbf{V}_{BA}\mathbf{R}_{\mathbf{\hat p}}.
\end{align}
Let
\begin{equation}\label{eq:R_Delta_hat_pE}
  \mathbf{R}_{\Delta\mathbf{\hat p}_E}\doteq\mathbf{R}_{\mathbf{\hat p}}-\mathbf{T}
\end{equation}
 which we see is the MSE matrix of the MMSE estimate $\mathbf{\hat p}_E$ of $\mathbf{\hat p}$ from $\mathbf{y}_{EA}$. Hence, $\mathbf{T} = \mathbf{R}_{\mathbf{\hat p}_E}\doteq\mathbb{E}\{\mathbf{\hat p}_E\mathbf{\hat p}_E^H\}$.
 %(Note $\lim_{p_A\to 0}\mathbf{R}_{\Delta\mathbf{\hat p}_E}=\mathbf{0}$.)

Then \eqref{eq:RDsE} becomes
\begin{align}
&\mathbf{R}_{\Delta\mathbf{s}_E}
                    =\mathbf{I}_{n_B}-\mathbf{H}_{EB}''^H\notag\\
  &\,\,\cdot
                    \left (\mathbf{H}_{EB}''(\mathbf{R}_{\Delta\mathbf{\hat p}_E}+\mathbf{I}_{n_B})\mathbf{H}_{EB}''^H
  +\mathbf{I}_{n_E}\right )^{-1}\mathbf{H}_{EB}''\notag\\
  &=\mathbf{I}_{n_B}-
                    \left (\mathbf{H}_{EB}''^H\mathbf{H}_{EB}''(\mathbf{R}_{\Delta\mathbf{\hat p}_E}+\mathbf{I}_{n_B})
  +\mathbf{I}_{n_B}\right )^{-1}\mathbf{H}_{EB}''^H\mathbf{H}_{EB}''\notag\\
  &=
                    \left (\mathbf{H}_{EB}''^H\mathbf{H}_{EB}''(\mathbf{R}_{\Delta\mathbf{\hat p}_E}+\mathbf{I}_{n_B})
  +\mathbf{I}_{n_B}\right )^{-1}\notag\\
  &\,\,\cdot
  \left (\mathbf{H}_{EB}''^H\mathbf{H}_{EB}''\mathbf{R}_{\Delta\mathbf{\hat p}_E}
  +\mathbf{I}_{n_B}\right ).
\end{align}
Hence the capacity of the effective return channel from Bob to Eve relative to $\mathbf{s}$ (in bits per round-trip symbol interval) is
\begin{align}\label{eq:CEB_MIMO}
  &C_{E|B,G}\doteq I(\mathbf{s};\{\mathbf{y}_{EA},\mathbf{y}_{EB}\})= I(\mathbf{s};\mathbf{\hat s}_E)=\log\frac{1}{|\mathbf{R}_{\Delta\mathbf{s}_E}|}\notag\\
  &=\log\frac{|\mathbf{H}_{EB}''^H
  \mathbf{H}_{EB}''(\mathbf{R}_{\Delta\mathbf{\hat p}_E}+\mathbf{I}_{n_B})
  +\mathbf{I}_{n_B}|}{|\mathbf{H}_{EB}''^H\mathbf{H}_{EB}''
  \mathbf{R}_{\Delta\mathbf{\hat p}_E}
  +\mathbf{I}_{n_B}|}\notag\\
  &\doteq \log\frac{N_{E|B}}{D_{E|B}}.
\end{align}
Again we have applied the jointly Gaussian nature of $\mathbf{s}, \mathbf{y}_{EA}, \mathbf{y}_{EB}$ for the 2nd and 3rd equalities in \eqref{eq:CEB_MIMO}.

\subsection{Secrecy rate of G-STEEP}
\begin{Theorem}\label{Theorem_MIMO_Cs_new} An achievable secrecy rate of G-STEEP  based on the effective wiretap-channel system from Bob to Alice against Eve (in bits per round-trip symbol interval or two complex channel uses) is
\begin{align}\label{eq:MIMO_Cs_new}
  &R_{s,G}\doteq (I(\mathbf{s};\{\mathbf{x}_A,\mathbf{y}_A\})
  -I(\mathbf{s};\{\mathbf{y}_{EA},\mathbf{y}_{EB}\}))^+\notag\\
  &=\left (C_{A|B,G}-C_{E|B,G}\right )^+=\left (\log\frac{N_{A|B}D_{E|B}}{D_{A|B}N_{E|B}}\right )^+\notag\\
  &=\left [\log\left (\frac{|\mathbf{H}_{AB}''^H\mathbf{H}_{AB}''
   (\mathbf{R}_{\Delta\mathbf{p}'}+\mathbf{I}_{n_B})+
  \mathbf{I}_{n_B}|}{|\mathbf{H}_{AB}''^H\mathbf{H}_{AB}''
  \mathbf{R}_{\Delta\mathbf{p}'}+
  \mathbf{I}_{n_B}|}\right .\right .\notag\\
  &\,\,\cdot\left .\left .\frac{|\mathbf{H}_{EB}''^H\mathbf{H}_{EB}''
  \mathbf{R}_{\Delta\mathbf{\hat p}_E}
  +\mathbf{I}_{n_B}|}{|\mathbf{H}_{EB}''^H\mathbf{H}_{EB}''
  (\mathbf{R}_{\Delta\mathbf{\hat p}_E}+\mathbf{I}_{n_B})
  +\mathbf{I}_{n_B}|}\right )\right ]^+.
\end{align}
where $\mathbf{R}_{\Delta\mathbf{p}'}$ is given in \eqref{eq:RDpP}, and $\mathbf{R}_{\Delta\mathbf{\hat p}_E}$ is given in \eqref{eq:R_Delta_hat_pE}.
\end{Theorem}
\begin{IEEEproof}
This follows from the WTC theory for Gaussian signaling over Gaussian noise channels \cite{Bloch2011} with respect to the message vector $\mathbf{s}$ from Bob, and the previous results shown in \eqref{eq:CAB_MIMO} and \eqref{eq:CEB_MIMO}.
\end{IEEEproof}

\subsection{Properties of secrecy rate of G-STEEP}\label{sec:STEEP_properties}
\begin{Lemma} Assuming constant channel matrices,
the secret-key capacity $C_{key}$ (in bits per probing symbol interval) based on the data sets at Alice, Bob and Eve after phase 1 of G-STEEP (and a public communication phase after that) is
\begin{align}\label{eq:MIMO_key}
  &C_{key}=\log\left |\mathbf{I}_{n_A}+\mathbf{H}_{BA}'^H\mathbf{H}_{BA}'
  \left (\mathbf{H}_{EA}'^H\mathbf{H}_{EA}'+\mathbf{I}_{n_A}\right )^{-1}\right |.
 % \notag\\
%  &=\log\left |\mathbf{I}_{n_B}+\right .\notag\\
%  &\,\,\left .\frac{p_A}{n_A\sigma_B^2}\mathbf{H}_{BA}
%  \left (\frac{p_A}{n_A\sigma_{EA}^2}\mathbf{H}_{EA}^H\mathbf{H}_{EA}+\mathbf{I}_{n_A}\right )^{-1}\mathbf{H}_{BA}^H\right |.
\end{align}
\end{Lemma}
\begin{IEEEproof}
This lemma is a special case of Theorem 1 in \cite{HuaMaksud2024} where Maurer's lower and upper bounds, generalized in \cite{Bloch2011} and further applied asymptotically to continuous sources via generalized mutual information, are used. (Some earlier attempt such as \cite{Naito2008} to improve Maurer's lower bound is not necessary here.) Specifically, the lower bound $I(\mathbf{x}_A;\mathbf{y}_B)-I(\mathbf{y}_B;\mathbf{y}_{E,A})$ of $C_{key}$, which reduces to $[h(\mathbf{y}_B)-
h(\mathbf{y}_B|\mathbf{x}_A)]
-[h(\mathbf{y}_B)-h(\mathbf{y}_B|\mathbf{y}_{E,A})]=h(\mathbf{y}_B|\mathbf{y}_{E,A})-
h(\mathbf{y}_B|\mathbf{x}_A)
$, equals the upper bound $I(\mathbf{x}_A;\mathbf{y}_B|\mathbf{y}_{E,A})
=h(\mathbf{y}_B|\mathbf{y}_{E,A})-h(\mathbf{y}_B|\mathbf{x}_A,\mathbf{y}_{E,A})
=h(\mathbf{y}_B|\mathbf{y}_{E,A})-h(\mathbf{y}_B|\mathbf{x}_A)$  where the last equation follows from the fact that $\mathbf{y}_B$ and $\mathbf{y}_{E,A}$ are independent of each other when conditioned on $\mathbf{x}_A$. One can verify that \eqref{eq:MIMO_key} follows from $C_{key}=h(\mathbf{y}_B|\mathbf{y}_{E,A})-h(\mathbf{y}_B|\mathbf{x}_A)$, which is also $\xi_B$ in equation (8) in \cite{HuaMaksud2024} with constant channel matrices, and $\mathbf{H}_{EA}$ and $\frac{p_A}{n_A}$ here are $\mathbf{G}_A$ and $\gamma_{BA}$ in \cite{HuaMaksud2024}. Furthermore,  $\lambda_B$ and $\lambda_{EA}$ in \cite{HuaMaksud2024} are both normalized to be one here. Alternatively, see \cite{RennaBloch2013}.
\end{IEEEproof}

Note that $C_{key}$ is the maximum secret-key rate achievable based on the data sets (among all possible statistical distributions of the probes) generated in phase 1 of G-STEEP at Alice, Bob and Eve and through communications in the public network to which Eve has full access. So, if  Eve's receive channel from Bob is no weaker than Alice's receive channel from Bob (in phase 2 of G-STEEP), we should expect $R_{s,G}\leq C_{key}$. A precise statement is shown later in Proposition \ref{Proposition_comparison}. If $R_{s,G}$ approaches $C_{key}$ under high powers, we can say that $R_{s,G}$ is optimal against strong Eve under high powers.

 But if Eve's receive channel from Bob is weaker than that at Alice, it is possible to have $R_{s,G}> C_{key}$. But one should not be excited by this situation. We know that $C_{key}$ is based on the assumption that all communications for secret key generation are done in the public domain. If any of these communications are not public (i.e., secure or partially secure), the resulting $C_{key}$ would be higher.  So, for a meaningful comparison between $R_{s,G}$ and $C_{key}$, we should assume that Eve's receive channel in phase 2 of G-STEEP is no weaker than that at Alice.

One may argue that the secret-key capacity $C_{key}'$ based on $\{\mathbf{x}_A,\mathbf{y}_A\}$ at Alice, $\{\mathbf{y}_B,\mathbf{s}\}$ at Bob and $\{\mathbf{y}_{EA},\mathbf{y}_{EB}\}$ at Eve after both phases of G-STEEP (and using additional and iterative operations for information reconciliation and privacy amplification via public communications) should always be larger than or equal to $R_{s,G}$. But the analysis of $C_{key}'$ is more involved, and there is a gap between its lower and upper bounds.

\begin{Proposition}\label{Proposition_DoF}
Assume that $\mathbf{H}_{AB}$, $\mathbf{H}_{BA}$, $\mathbf{H}_{EA}$ and $\mathbf{H}_{EB}$ are typical realizations (where the rank of each matrix equals to the minimum of its numbers of rows and columns and the rank conditions in \eqref{eq:rank_cond_1}, \eqref{eq:rank_cond_2} and \eqref{eq:rank_cond_3} hold). For $n_A\geq n_B$, $n_E\geq 1$ and any given (fixed) $\eta_p =\frac{p_B}{p_A}$,
\begin{align}\label{}
  &\lim_{p_A\to\infty} \frac{1}{\log p_A}R_{s,G}=\lim_{p_A\to\infty}\frac{1}{\log p_A}C_{key}\notag\\
  &=\min(n_B,(n_A-n_E)^+),
\end{align}
i.e., $\texttt{DoF}(R_{s,G})=\texttt{DoF}(C_{key})$. Namely, $R_{s,G}$ is optimal in DoF.
\end{Proposition}

\begin{IEEEproof}
See Appendix\ref{sec:Proof_Proposition_DoF}.
\end{IEEEproof}

The above DoF is the maximum achievable DoF currently known, which is consistent with a prior result in \cite{Alouini2016}.

The DoF only depends on the numbers of antennas on Alice, Bob and Eve, which is not affected by any finite scaling on channel matrices and/or on noise variances.

\begin{Proposition}\label{Proposition_Ckey_Cs}
Assume typical realizations of all channel matrices (like those in Proposition \ref{Proposition_DoF}).
  For $n_E\geq n_A\geq n_B\geq 1$,
  \begin{align}\label{eq:MIMO_optimal}
    &\lim_{p_A\to\infty}(\lim_{p_B\to\infty} R_{s,G})=\lim_{p_A\to\infty} C_{key}\notag\\
    &=\log\left |\mathbf{I}_{n_B}+\boldsymbol{\Pi}_{BA}^2\mathbf{V}_{BA}^H
  (\mathbf{H}_{EA}^H\mathbf{H}_{EA})^{-1}\mathbf{V}_{BA}\right |.
  \end{align}
  Namely, $R_{s,G}$ is optimal (against strong Eve) asymptotically as $p_A\to\infty$ and $\frac{p_B}{p_A}\to\infty$.
\end{Proposition}
\begin{IEEEproof}
See Appendix\ref{sec:Proof_Proposition_Ckey_Cs}.
\end{IEEEproof}

The above proposition is also intuitively justified if we think of $\frac{p_B}{p_A}\to\infty$ as somewhat similar to the case where phase 2 of G-STEEP only uses public communications and also think of $p_A\to\infty$ as somewhat similar to the case where the encryption in phase 2 is done via a modulo sum between two discrete random variables. In other words, for $\frac{p_B}{p_A}\to\infty$, both Alice and Eve would receive the same $\sqrt{p_B}(\mathbf{\hat p}+\mathbf{s})$ from Bob, i.e., the phase 2 would be via a public channel. For $p_B\to\infty$, $\sqrt{p_B}(\mathbf{\hat p}+\mathbf{s})=\sqrt{p_B}\mathbf{\hat p}+\sqrt{p_B}\mathbf{s}$ is a sum between $\sqrt{p_B}\mathbf{\hat p}$ and $\sqrt{p_B}\mathbf{s}$ (which would be virtually uniformly distributed), and this sum would be like a modulo-sum with an infinite modulo. Then the encryption lemma would suggest that in the case of $p_A\to\infty$ and $\frac{p_B}{p_A}\to\infty$, $C_{key}=R_{s,G}$.  Again, the above discussion is no proof but only an intuition to make an intuitive sense of the result actually proven in Appendix\ref{sec:Proof_Proposition_Ckey_Cs}.

Since the limit in \eqref{eq:MIMO_optimal} is always positive (unless $\mathbf{H}_{EA}$ has an infinite norm), this proposition also suggests that for a sufficiently large (but finite) $p_A$ and a sufficiently large (but finite) $\frac{p_B}{p_A}$, $R_{s,G}$ is positive. We will see a more specific case of this next.

\begin{Corollary}\label{Corollary_1}
If $n_B=1$ and $\mathbf{H}_{AB}$, $\mathbf{H}_{BA}$ and $\mathbf{H}_{EB}$ are replaced by $\mathbf{h}_{AB}$, $\mathbf{h}_{BA}$ and $\mathbf{h}_{EB}$ (similarly for their scaled versions), then $R_{s,G}\doteq\left (C_{A|B,G}-C_{E|B,G}\right )^+$ with
\begin{align}\label{eq:CAB}
  &C_{A|B,G}=\log\left (1+\frac{\frac{S_{AB}}{2}}{\frac{1}{2}\frac{S_{AB}S_{BA}}{(S_{BA}+1)^2}+1}\right ),
\end{align}
\begin{align}\label{eq:CEB}
  &C_{E|B,G}=\log\left (1+\frac{\frac{S_{EB}}{2}}{(\sigma_{\hat p_0}^2-t)\frac{S_{EB}}{2}+1}\right ),
\end{align}
where $S_{AB}=\|\mathbf{h}_{AB}'\|^2$, $S_{BA}=\|\mathbf{h}_{BA}'\|^2$, $S_{EB}=\|\mathbf{h}_{EB}'\|^2$, $\sigma^2_{\hat p_0} = \frac{S_{BA}}{S_{BA}+1}$ and
\begin{align}\label{eq:t}
&t=\mathbf{r}^H\mathbf{H}_{EA}'^H
\left (\mathbf{H}_{EA}'\mathbf{H}_{EA}'^H+\mathbf{I}_{n_E}\right )^{-1}
\mathbf{H}_{EA}'\mathbf{r}
%\notag\\
%&=\frac{p_A\sigma_{\hat p_0}^4}{n_A}\mathbf{h}_{EA}^H
%\left (\frac{p_A}{n_A}\mathbf{h}_{EA}\mathbf{h}_{EA}^H+
%\frac{p_A}{n_A}\mathbf{G}_1\mathbf{G}_1^H+\sigma_{EA}^2\mathbf{I}_{n_E}\right )^{-1}
%\mathbf{h}_{EA}.
\end{align}
with $\mathbf{r}=\sigma^2_{\hat p_0}\frac{1}{\|\mathbf{h}_{BA}\|}\mathbf{h}_{BA}^*$.
\end{Corollary}
\begin{IEEEproof}
This follows from Theorem \ref{Theorem_MIMO_Cs_new}. In particular, $\mathbf{T}$ in \eqref{eq:T} is now reduced to the scalar $t$.
\end{IEEEproof}

We see that for the case of $n_B=1$, the effects of $\mathbf{h}_{AB}'$, $\mathbf{h}_{BA}'$ and $\mathbf{h}_{EB}'$ on $R_{s,G}$ are only through their norms. The effect of $\mathbf{H}_{EA}'$ on $R_{s,G}$ is only through the scalar $t$.

\subsection{The special case of $n_A=n_B=1$}\label{sec:nAnB1}

\subsubsection{Analysis of $R_{s,G}$}

For $n_A=n_B=1$, we let $\mathbf{H}_{AB}$, $\mathbf{H}_{BA}$, $\mathbf{H}_{EA}$ and $\mathbf{H}_{EB}$ be replaced by $h_{AB}$, $h_{BA}$, $\mathbf{h}_{EA}$ and $\mathbf{h}_{EB}$.
Then
it follows from Corollary \ref{Corollary_1} that
\begin{align}\label{eq:R_s_new}
  &R_{s,G}=\left [\log\left (1+\frac{b/2}{bA_1/2+1}\right )-\log\left (1+\frac{\beta b/2}{\beta bA_2/2 +1}\right )\right ]^+
\end{align}
with $a\doteq S_{BA}\doteq p_A|h_{BA}|^2=|h_{BA}'|^2$ and $b\doteq S_{AB}\doteq p_B|h_{AB}|^2=|h_{AB}'|^2$. Also $A_1=\frac{a}{(a+1)^2}$ and $A_2=A_1\frac{(a+\alpha a+1)}{\alpha a+1}$ with  $\alpha\doteq\frac{S_{EA}}{S_{BA}}$ and $\beta\doteq\frac{S_{EB}}{S_{AB}}$. Furthermore, $S_{EA}\doteq p_A \|\mathbf{h}_{EA}\|^2=\|\mathbf{h}_{EA}'\|^2$ and $S_{EB}\doteq p_B \|\mathbf{h}_{EB}\|^2=\|\mathbf{h}_{EB}'\|^2$.
Note that $A_1<A_2<1$ and they are invariant to $b$.

In this special case, all channel gains and noise variances are completely lumped into just four parameters: $a$, $b$, $\alpha$ and $\beta$. Here $a$ and $b$ are respectively the (raw channel) SNR at Bob in phase 1 and the (raw channel) SNR at Alice in phase 2. And $a$ and $b$ are proportional to $p_A$ and $p_B$ respectively. Furthermore, $\alpha$ and $\beta $ are the SNR ratios measuring Eve's (raw) channel strengths over users' (raw) channel strengths in phases 1 and 2 respectively. It is important to distinguish between ``raw channels'' and ``effective channels'', the latter of which are induced by STEEP.

In particular, if Eve's (raw) channel is stronger than users' (raw) channel in phase 1, then $\alpha>1$; and if Eve's (raw) channel is stronger than users' (raw) channel in phase 2, then $\beta>1$.

If $\alpha\geq 1$ and $\beta\geq 1$, all conventional WTC schemes either from Alice to Bob or from Bob to Alice have zero secrecy capacity.

\begin{Proposition}\label{Proposition_positive_Cs_nA_1}
For $n_A=n_B=1$, $R_{s,G}>0$ if and only if
\begin{align}\label{eq:cond1}
  &b>\bar b\doteq\frac{2(\beta-1)}{\beta (A_2-A_1)}=\frac{2(\beta-1)(a+1)^2(\alpha a+1)}{\beta a^2}.
\end{align}
 %Also  for $\beta>1$, $R_{s,G}$ increases as $b$ increases.
\end{Proposition}
\begin{IEEEproof}
This can be directly verified from \eqref{eq:R_s_new}.
\end{IEEEproof}

We see that as $a$ (or equivalently $p_A$) either decreases to zero or increases to infinity,  $\bar b$ increases to infinity subject to $\beta>1$. But for $\alpha>1$, $\beta>1$ and $a\gg1$, we have
\begin{equation}\label{}
  \bar b \approx 2\frac{\beta-1}{\beta}\alpha a =\mathcal{O}(\alpha a)
\end{equation}

%Hence, for $\alpha>1$, $\beta>1$ and a given $b$ (such as 20dB to 30dB in practice), the optimal value of $a$ to maximize $R_{s,G}$ is generally in between zero and $b$.

In practice, one can utilize \eqref{eq:cond1} to ensure a positive secrecy rate whenever an upper bound on $\alpha$ (not necessarily on $\beta$) is available. In the case of random fading channels, the probability for \eqref{eq:cond1} not to hold can be kept small by keeping a large ratio of $p_B$ over $p_A$ \cite{HuaICC2024}.

One can also verify that
$R_{s,G}$ increases as $\alpha$ and/or $\beta$ decrease; for $\beta>1$, $R_{s,G}$ increases as $b$ increases, but $R_{s,G}$ saturates as $b$ becomes large; and $R_{s,G}$ versus $a$ is not monotonic in general. For a given $b$, $R_{s,G}$ generally peaks at a value of $a$ in between zero and $b$.

%Numerical illustrations of $R_{s,G}$ for fading channels are shown in \cite{HuaICC2024}.

\subsubsection{Comparison to $C_{key}$}
For $n_A=n_B=1$, \eqref{eq:MIMO_key} reduces to
\begin{align}\label{eq:SISO_key}
  &C_{key} = \log\left (1+\frac{S_{BA}}{S_{EA}+1}\right)
  =\log\left (1+\frac{a}{\alpha a+1}\right)=\log \frac{A_2}{A_1}.
\end{align}

\begin{Proposition}\label{Proposition_comparison}
  For any given $\alpha$ and $a$, there is a sufficiently large (but finite) $b$ and a sufficiently small (positive) $\beta$ such that $C_{key}<R_{s,G}$. But if $\beta\geq 1$, then $C_{key}>R_{s,G}$ for any finite $\alpha$, $a$ and $b$.
\end{Proposition}
\begin{IEEEproof}
Let $\gamma>1$ be such that $\gamma A_2-A_1<1$, i.e., $\gamma<\frac{1-A_1}{A_2}$.
The first term of $R_{s,G}$ in \eqref{eq:R_s_new} is strictly larger than $C_{key}+\log\gamma$ if  $\frac{b}{2}>\frac{\gamma\frac{A_2}{A_1}-1}{1-(\gamma A_2-A_1)}$. The second term of $R_{s,G}$ in \eqref{eq:R_s_new} is smaller than $\log\gamma$ if
$
  \beta<\frac{\gamma-1}{\frac{b}{2}(1-(\gamma-1)A_2)}
$.
This proves the first statement in the proposition. To prove the second statement, first consider $\beta=1$. In this case, ``$C_{key}>R_{s,G}$'' is equivalent to
\begin{equation}\label{eq:inequality}
  \frac{A_2}{A_1}-1>\frac{\frac{b}{2}\left (1-\frac{A_2}{A_1}\right )}{\left (\frac{b}{2}A_1+1\right )\left (\frac{b}{2}A_2+1\right )}
\end{equation}
which always holds since the left side of \eqref{eq:inequality} is positive and the right side of \eqref{eq:inequality} is negative. Finally, notice that $R_{s,G}$ is a decreasing function of $\beta$.
\end{IEEEproof}

Alternatively, it follows from \eqref{eq:R_s_new} that
\begin{align}\label{}
  &\lim_{b\to\infty} R_{s,G}=\left (\log\left (1+\frac{1}{A_1}\right )
  -\log\left (1+\frac{1}{A_2}\right )\right )^+\notag\\
  &=\left ( \log\frac{A_2(A_1+1)}{A_1(A_2+1)}\right )^+
  <C_{key}
\end{align}
Since $R_{s,G}$ increases with $b$ for $\beta>1$, then for $\beta>1$ we have $R_{s,G}<C_{key}$ for all $\alpha$, $a$ and $b$, which is consistent with Proposition \ref{Proposition_comparison}.

However,
\begin{equation}\label{eq:a1}
  \lim_{a\to\infty} C_{key}=\log\left (1+\frac{1}{\alpha}\right ),
\end{equation}
which is the same as
\begin{align}\label{eq:a2}
  &\lim_{a\to\infty}(\lim_{b\to\infty}R_{s,G})=\lim_{a\to\infty}\left ( \log\frac{A_2(A_1+1)}{A_1(A_2+1)}\right )^+\notag\\
  &=
  \log\left (1+\frac{1}{\alpha}\right ).
\end{align}
In a practical term, we can say that if both $a$ and $b$ are  large while $b$  dominates $a$, then $R_{s,G}\approx C_{key}$. This is a special case of Proposition \ref{Proposition_Ckey_Cs}.

\section{STEEP with PSK channel Probing and PSK Nonlinear Encryption (P-STEEP)}\label{sec:P-STEEP}
In this section, P-STEEP is presented assuming  $n_A=n_B=1$ and $n_E\geq 1$.
It is important to note that for applications where power control is difficult (due to nonlinearity of power amplifier, channel disturbances, etc), nonlinear modulation such as PSK is always preferred to linear modulation.
%A key difference between P-STEEP and G-STEEP is how the encryption is done by Bob on the estimated probes before they are echoed back.

\subsection{Description of P-STEEP}
In phase 1 of P-STEEP, Alice sends out PSK probes $\sqrt{p_A}x_A=\sqrt{p_A}e^{j\theta}$ where $\theta$ is an M-ary discrete uniform random variable within $[-\pi,\pi]$. Then Bob receives
\begin{equation}\label{}
  y_B = \sqrt{p_A}h_{BA} x_A+w_B.
\end{equation}
A sufficient statistic from $y_B$ for $x_A=e^{j\theta}$ (at Bob) is
\begin{equation}\label{}
  r_B\doteq\frac{1}{\sqrt{p_A}h_{BA}}y_B=x_A+v_B
\end{equation}
where $v_B$ is $\mathcal{CN}(0,\frac{1}{S_{BA}})$ with $S_{BA}=p_A|h_{BA}|^2$.

In phase 2 of P-STEEP, Bob applies PSK nonlinear encryption, i.e., he sends out $\sqrt{p_B}x_B = \sqrt{p_B}e^{j\phi}r_B$ where $\phi$ is a secret phase value (meant for Alice) randomly chosen (in this paper) from the same discrete constellation as $\theta$. Here the construction of $x_B=e^{j\phi}r_B$ is different from that for G-STEEP with $n_A=n_B=1$. This nonlinear encryption fits naturally with PSK (a nonlinear modulation).

It is important to note that while both $\theta$ and $\phi$ are discrete, $r_B$ here is continuous. The use of continuous $r_B$ (instead of a quantized $r_B$ with the constellation size $M$) to construct $x_B=e^{j\phi}r_B$ reduces the computational complexity at Bob (i.e., no detection is needed at Bob). It is however not clear whether this would yield a better secrecy rate than the quantized option. There is also a strategy in between ``completely hard'' and ``completely soft'', i.e., replacing $r_B$ by its quantized value with a constellation size equal to $lM$ with $l\geq 1$. When $l=1$, we say that the quantized $r_B$ is completely hard. As $l$ becomes larger, the quantized $r_B$ becomes ``softer''. But in this paper, we only focus on continuous $r_B$.

\subsection{Analysis of the signal received by Alice in phase 2}
The signal received by Alice in phase 2 is
\begin{equation}\label{eq:yA2}
  y_A = \sqrt{p_B}h_{AB}x_B+w_A.
\end{equation}
A sufficient statistic from $y_A$ for $\phi$ (at Alice) is
\begin{equation}\label{eq:rA}
  r_A\doteq\frac{x_A^*}{\sqrt{p_B}h_{AB}}y_A = e^{j\phi}+e^{j\phi}x_A^*v_B+x_A^*v_A
\end{equation}
where  $v_A$ is $\mathcal{CN}(0,\frac{1}{S_{AB}})$ with $S_{AB}=p_B|h_{AB}|^2$.

\begin{Lemma}\label{Lemma_Gaussian}
If $A$ is a circular complex Gaussian random variable with zero mean and variance $\sigma^2$, i.e., $\mathcal{CN}(0,\sigma^2)$, then so is $e^{j\theta}A$ for any $\theta$. If $A$ and $B$ are two independent  circular complex Gaussian random variables with zero means and variances $\sigma_A^2$ and $\sigma_B^2$ respectively, then so are $e^{j\theta_A}A$ and  $e^{j\theta_B}B$ for any $\theta_A$ and $\theta_B$.
\end{Lemma}
\begin{IEEEproof}
Since $A$ is $\mathcal{CN}(0,\sigma^2)$, its amplitude $|A|$ and phase $\angle A$ are independent variables with $|A|$ being Rayleigh distributed and $\angle A$ being uniform distributed within $[0,2\pi)$. For any $\theta$, the modulo sum $(\theta+\angle A)_{mod-2\pi}$ remains uniform with $[0,2\pi)$. Hence the distributions of $|e^{j\theta}A|=|A|$ and $\angle (e^{j\theta}A)=(\theta+\angle A)_{mod-2\pi}$ do not change with $\theta$, i.e., $e^{j\theta}A$ is also $\mathcal{CN}(0,\sigma^2)$. The other statement can be similarly proved (even if $\theta_A=\theta_B$).
\end{IEEEproof}

Since $v_B$ and $v_A$ are independent circular complex Gaussian, we can also write
\begin{equation}\label{}
  r_A\doteq\frac{x_A^*}{\sqrt{p_B}h_{AB}}y_A = e^{j\phi}+v_B'+v_A'
\end{equation}
where $v_B'$ and $v_A'$ are independent of $\phi$ and each other, and they have the same distributions as $v_B$ and $v_A$.

The minimum distance between the constellation points of $e^{j\phi}$ is $2\sin\frac{\pi}{M}$. Hence the error rate in detecting  $e^{j\phi}$ from $r_A$ is (approximately for $M=2^m$ with $m\geq 2$)
\begin{equation}\label{}
  p_{e,A} = n_0Q\left (\frac{\sin\frac{\pi}{M}}{\epsilon_A}\right )
\end{equation}
where $n_0=1$ for $m=1$, $n_0=2$ for $m\geq 2$,
\begin{equation}\label{eq:eA}
  \epsilon_A=\sqrt{\frac{1}{2S_{BA}}+\frac{1}{2S_{AB}}}
\end{equation}
and $Q(x)=\int_x^\infty \frac{1}{\sqrt{2\pi}}e^{-u^2/2}du$. With Gray mapping of bits, $p_{e,A}$ is also the (uncoded) secret-bit error rate suffered by Alice for all $m\geq 1$.

The effective capacity from Bob to Alice relative to $\phi$ is
\begin{align}\label{}
  &C_{A|B,P} \doteq I(\phi;r_A)=H(\phi)-H(\phi |r_A),
\end{align}
where $H(\phi)=\log M$ (the entropy of $\phi$).
To determine $H(\phi |r_A)$, we can view $\phi$ given $r_A$ as the optimal decision (also known as hard decision) of $\phi$ from $r_A$.

For $M=2$, the optimal decision of $\phi$ from $r_A$ takes two possible values with the probabilities $1-p_{e,A}$ and $p_{e,A}$ respectively. In this case,
\begin{align}\label{eq:M2}
  &C_{A|B,P} = 1-h_2(p_{e,A})
\end{align}
with $h_2(p)\doteq-p\log p -(1-p)\log (1-p)$.

For $M=2^m$ with $m\geq 2$, the optimal decision of $\phi$ from $r_A$ at a high SNR (i.e., small $p_{e,A}$) takes approximately three possible values: the correct $\phi$ with the probability $1-p_{e,A}$, and the two nearest neighbors of $\phi$ with the probability $\frac{1}{2}p_{e,A}$ for each. In this case, we can write
\begin{align}\label{eq:M4}
  &C_{A|B,P}\approx m -2\left (\frac{1}{2}p_{e,A}\log\frac{1}{\frac{1}{2}p_{e,A}} \right )\notag\\
  &\,\,-(1-p_{e,A})\log\frac{1}{1-p_{e,A}}\notag\\
  &=m-p_{e,A}-h_2(p_{e,A})
\end{align}
with  $m\geq 2$.

\subsection{Analysis of the signals received by Eve in phases 1 and 2}
After phases 1 and 2, the signals received by Eve are
\begin{align}
&\mathbf{y}_{EA} =\sqrt{p_A}\mathbf{h}_{EA}x_A+\mathbf{w}_{EA},\\
&\mathbf{y}_{EB} =\sqrt{p_B}\mathbf{h}_{EB}x_B+\mathbf{w}_{EB},
\end{align}
or equivalently
\begin{align}
&r_{EA} \doteq\frac{\mathbf{h}_{EA}^H}{\sqrt{p_A}\|\mathbf{h}_{EA}\|^2}\mathbf{y}_{EA}=x_A+v_{EA},\\
&r_{EB} \doteq\frac{\mathbf{h}_{EB}^H}{\sqrt{p_A}\|\mathbf{h}_{EB}\|^2}\mathbf{y}_{EB}=x_B+v_{EB}.
\end{align}
Here $v_{EA}$ is $\mathcal{CN}(0,\frac{1}{S_{EA}})$ with $S_{EA}=p_A\|\mathbf{h}_{EA}\|^2$, and $v_{EB}$ is $\mathcal{CN}(0,\frac{1}{S_{EB}})$ with $S_{EB}=p_B\|\mathbf{h}_{EB}\|^2$.

Let us consider
\begin{equation}\label{}
  r_E\doteq r_{EA}^*r_{EB} = e^{j\phi}+x_A^*e^{j\phi}v_B+x_A^*v_{EB}+v_{EA}^*e^{j\phi}x_A
\end{equation}
where we have ignored the second-order terms of noises:  $e^{j\phi}v_{EA}^*v_B$ and $v_{EA}^*v_{EB}$. Since $v_B$, $v_{EB}$ and $v_{EA}$ are independent circular complex Gaussian, we can also write
\begin{equation}\label{}
  r_E\doteq r_{EA}^*r_{EB} = e^{j\phi}+v_B'+v_{EB}'+v_{EA}'
\end{equation}
where $v_B'$, $v_{EB}'$ and $v_{EA}'$ are also independent circular complex Gaussian and are independent of $\phi$ and $x_A$, and they have the same distributions as $v_B$, $v_{EB}$ and $v_{EA}$ respectively.

Since $\{r_{EA}, r_{EB}\}$ is a one-to-one function of $\{r_{EA}, r_E\}$, and $r_{EA}$ is approximately independent of $r_E$ and $\phi$, we now know that $r_E$ is a sufficient statistic from $\{r_{EA}, r_{EB}\}$ for $\phi$.

So the optimal detection of $e^{j\phi}$ from $\{r_{EA}, r_{EB}\}$ is the same as that from $r_E$. We know that the error rate in detecting $e^{j\phi}$ from $r_E$ is (approximately for $M=2^m\geq 4$)
\begin{equation}\label{}
  p_{e,E} = n_0Q\left (\frac{\sin\frac{\pi}{M}}{\epsilon_E}\right )
\end{equation}
where
\begin{equation}\label{eq:eE}
  \epsilon_E = \sqrt{\frac{1}{2S_{BA}}+\frac{1}{2S_{EA}}+\frac{1}{2S_{EB}}}.
\end{equation}

Similar to $C_{A|B,P}$ in \eqref{eq:M2} and \eqref{eq:M4}, we can express the effective capacity $C_{E|B,P}$ from Bob to Eve relative to $\phi$  as
\begin{equation}\label{}
  C_{E|B,P}=m-p_{e,E}-h_2(p_{e,E}).
\end{equation}

\subsection{Achievable secrecy rate}
An achievable secrecy rate of P-STEEP is
\begin{equation}\label{}
  R_{s,P}\doteq(C_{A|B,P}-C_{E|B,P})^+
=h_2(p_{e,E})-h_2(p_{e,A})
\end{equation}
which is positive if and only if $p_{e,A}<p_{e,E}$.
We see that $p_{e,A}<p_{e,E}$ if and only if $\epsilon_A<\epsilon_E$. It follows from
\eqref{eq:eA} and \eqref{eq:eE} that
\begin{equation}\label{}
  \frac{\epsilon_A^2}{\epsilon_E^2}=\frac{\frac{1}{a}+\frac{1}{b}}{\left (1+\frac{1}{\alpha}\right )\frac{1}{a}+\frac{1}{\beta b}},
\end{equation}
where $a=S_{BA}$, $b=S_{AB}$, $\alpha=\frac{S_{EA}}{S_{BA}}$ and $\beta=\frac{S_{EB}}{S_{AB}}$.
Hence $\epsilon_A<\epsilon_E$ if and only if
\begin{equation}\label{eq:power_condition}
  \frac{b}{a}>\alpha \left (1-\frac{1}{\beta}\right ).
\end{equation}
 The condition \eqref{eq:power_condition} always holds if $\beta<1$ (i.e., Eve's receive channel from Bob is weaker than Alice's receive channel from Bob). Otherwise, for $\beta>1$, the condition \eqref{eq:power_condition} can be met by a sufficiently large but finite $p_B$ while $p_A$ is finite (subject to all other parameters being finite).

\subsection{Ratio of bit error rates}
For large $a$ and $b$, both $p_{e,A}$ and $p_{e,E}$ are small subject to $\alpha>1$ and $\beta>1$. In this case, $R_{s,P}$ is only a small positive value under \eqref{eq:power_condition}. But the ratio $\gamma_p$ of $p_{e,A}$ over $p_{e,E}$ is also a meaningful metrics subject to a sufficiently long packet (e.g., a packet of $n$ independent bits with $np_{e,E}\approx 1$).

Applying $\frac{x}{1+x^2}\phi_0(x)<Q(x)<\frac{\phi_0(x)}{x}$ with $\phi_0(x)=\frac{1}{\sqrt{2\pi}}e^{-\frac{x^2}{2}}$ and the condition \eqref{eq:power_condition}, one can verify that
\begin{align}
&\gamma_p \doteq \frac{p_{e,A}}{p_{e,E}}<(1+\delta_p)\exp\left (-P\right )
\end{align}
where
\begin{equation}\label{}
  \delta_p = \frac{\epsilon_A\epsilon_E}{\sin^2\frac{\pi}{M}},
\end{equation}
\begin{equation}\label{}
  P=\sin^2\left (\frac{\pi}{M}\right) \frac{a^2b(\beta b +\alpha a -\alpha \beta a)}{(a+b)(\alpha\beta ab+\beta ab +\alpha a^2)}.
\end{equation}
Here $1+\delta_p\approx 1$ for large $a$ and $b$. To obtain a large $P$ and hence a very small $\gamma_p$, we need a large $a$ and a large $\frac{b}{a}$ because
\begin{equation}\label{}
  \lim_{b\to\infty} P= \sin^2\left (\frac{\pi}{M}\right)\frac{a}{\alpha +1}
\end{equation}
which increases with $a$.
For example, if $M=2$, $\alpha=\beta=2$, $a=10^2$ and $b=10^3$ (i.e., 20dB and 30dB respectively), we have $P\approx26.4$.

\section{STEEP for Multiple Access (M-STEEP)}\label{sec:M-STEEP}
Let us now go back to G-STEEP but consider its use for multiple access.
Specifically, let there be an access point (AP) with $n_A$ antennas, and $M$ units of single-antenna user equipment (UE) which are denoted by UE$_1$, $\cdots$, UE$_M$. If we apply G-STEEP to AP and each UE separately, there would be a significant overhead associated with the channel probing for each UE. To reduce the overhead, an option is to allow all UEs to take advantage of the same probes transmitted by the AP. We will show a power condition under which the secrecy rate from each UE to AP stays positive for any given $M$.

\subsection{Description of M-STEEP}
In phase 1 of M-STEEP, AP broadcasts a sequence of independent realizations of the random probing vector $\sqrt{p_A/n_A}\mathbf{x}\in\mathbb{C}^{n_A\times 1}$ with $\mathbf{x}$ being $\mathcal{CN}(\mathbf{0},\mathbf{I}_{n_A})$. Then UE$_i$ receives
\begin{equation}\label{}
  y_i=\sqrt{p_A/n_A}\mathbf{h}_i^T\mathbf{x}+w_i=\mathbf{h}_i'^T\mathbf{x}+w_i
\end{equation}
with $i=1,\cdots,M$, $\mathbf{h}_i'=\sqrt{p_A/n_A}\mathbf{h}_i$ and $w_i$ being $\mathcal{CN}(0,1)$.  The effective probe arriving at UE$_i$ is defined to be $p_i=\mathbf{\bar h}_i^T\mathbf{x}$ with $\mathbf{\bar h}_i^T\doteq\frac{\mathbf{h}_i^T}{\|\mathbf{h}_i\|}$. The MMSE estimate of $p_i$ is denoted by $\hat p_i$, and its MSE is
\begin{equation}\label{}
  d_i\doteq\frac{1}{S_i+1}
\end{equation}
with $S_i=(p_A/n_A)\|\mathbf{h}_i\|^2$. The variance of $\hat p_i$ is
\begin{equation}\label{}
  c_i\doteq 1-d_i=\frac{S_i}{S_i+1}.
\end{equation}
 One can also verify that
$\mathbb{E}\{p_ip_j^*\}=\phi_{i,j}$, $\mathbb{E}\{\hat p_i\hat p_j^*\}=c_ic_j\phi_{i,j}$, and $\mathbb{E}\{p_i\hat p_j^*\}=c_j\phi_{i,j}$
 with $\phi_{i,j}=\mathbf{\bar h}_i^T\mathbf{\bar h}_j^*$.

In phase 2 of M-STEEP, the UEs use orthogonal multiple access to the AP. Specifically, UE$_i$ sends out a sequence of random realizations of $\sqrt{p_{ui}/2}(\hat p_i +s_i)$ (of power upper bounded by $p_{ui}$) with $s_i$ being a secret random symbol with the distribution $\mathcal{CN}(0,1)$, and the corresponding signal received by the AP is
\begin{align}\label{}
  &\mathbf{y}_{Ai} = \sqrt{p_{ui}/2}(\hat p_i +s_i)\mathbf{h}_{Ai}+\mathbf{w}_{Ai}\notag\\
  &=\sqrt{1/2}(\hat p_i +s_i)\mathbf{h}_{Ai}'+\mathbf{w}_{Ai}\in\mathbb{C}^{n_A\times 1}
\end{align}
with $\mathbf{w}_{Ai}$ being $\mathcal{CN}(\mathbf{0},\mathbf{I})$ and $\mathbf{h}_{Ai}'=\sqrt{p_{ui}}\mathbf{h}_{Ai}$.

\subsection{Effective channel from each UE to AP}
It follows from \eqref{eq:CAB_MIMO} with $n_B=1$ or from \eqref{eq:CAB} that an achievable rate from UE$_i$ to AP relative to $s_i$ (i.e., AP uses the signal from UE$_i$ and the original probing vector to extract the information from $s_i$) is
\begin{equation}\label{eq:RAi}
   R_{A|i} \doteq I(s_i;\hat s_i)=
   \log\left (1+\frac{\frac{S_{Ai}}{2}}{\frac{S_iS_{Ai}/2}{(S_i+1)^2}+1}\right ),
 \end{equation}
 where $\hat s_i$ is the MMSE estimate of $s_i$ by AP from $\mathbf{y}_{Ai}$ and $\mathbf{x}$, $S_{Ai}=p_{ui}\|\mathbf{h}_{Ai}\|^2$ and $S_i$ was defined before.

 Note that when $M\geq 2$, $R_{A|i}$ is a lower bound on $C_{A|i}\doteq I(s_i;\mathbf{y}_A|\mathbf{x})$ with $\mathbf{y}_A=[\mathbf{y}_{A1}^T,\cdots,\mathbf{y}_{AM}^T]^T$, the latter of which represents the optimal effective channel capacity from UE$_i$ to AP. However, even with $R_{A|i}$, we still can show that M-STEEP achieves a positive secrecy rate for each UE.

 \subsection{Effective channel from each UE to Eve}
 The signals received by Eve during both phases of M-STEEP are
 \begin{equation}\label{eq:yEA}
   \mathbf{y}_{EA} =\sqrt{p_A/n_A}\mathbf{H}_{EA}\mathbf{x}+\mathbf{w}_{EA}= \mathbf{H}_{EA}'\mathbf{x}+\mathbf{w}_{EA},
 \end{equation}
 \begin{align}\label{eq:yEi}
   &\mathbf{y}_{Ei} =\sqrt{p_{ui}/2}\mathbf{h}_{Ei}(\hat p_i+s_i)+\mathbf{w}_{Ei}\notag\\
   &= \sqrt{1/2}\mathbf{h}_{Ei}'(\hat p_i+s_i)+\mathbf{w}_{Ei},
 \end{align}
 for all $i=1,\cdots,M$. Here $\mathbf{h}_{Ei}'=\sqrt{p_{ui}}\mathbf{h}_{Ei}$, and $\hat p_i$ for every $i$ depends on $\mathbf{x}$. Also note that $s_1,\cdots,s_M$ are independent of each other.

 It can be shown that the MSE of the MMSE estimate $\hat s_{iE}$ of $s_i$ by Eve using $\mathbf{y}_E\doteq[\mathbf{y}_{E1}^T,\cdots,\mathbf{y}_{EM}^T,\mathbf{y}_{EA}^T]^T$ is
 \begin{equation}\label{eq:s_D_s}
   \sigma_{\Delta s_{iE}}^2 =
   1-\mathbf{r}_i^H\mathbf{R}^{-1}\mathbf{r}_i
 \end{equation}
 where $\mathbf{r}_i^H =\mathbb{E}\{s_i\mathbf{y}_E^H\}$ and $\mathbf{R}=\mathbb{E}\{\mathbf{y}_E\mathbf{y}_E^H\}$. Furthermore,
 \begin{equation}\label{eq:ri}
   \mathbf{r}_i = \left [\mathbf{0}_{n_E(i-1)}^T,\sqrt{1/2}\mathbf{h}_{Ei}'^T,\mathbf{0}_{n_E (M-i+1)}^T \right ]^T
 \end{equation}
 and
 \begin{equation}\label{eq:R}
   \mathbf{R} = \left [\begin{array}{ccc}
                         \mathbf{R}_{1,1} & \cdots  & \mathbf{R}_{1,M+1} \\
                         \cdots & \cdots & \cdots  \\
                         \mathbf{R}_{M+1,1} & \cdots  & \mathbf{R}_{M+1,M+1}
                       \end{array}
    \right ].
 \end{equation}
Here $\mathbf{0}_m$ is a zero vector of $m$ elements, and $\mathbf{R}_{i,j}=\mathbf{R}_{j,i}^H$ for all $i$ and $j$. For $1\leq i\leq M$, $1\leq j\leq M$ and $i\neq j$,
\begin{equation}\label{}
  \mathbf{R}_{i,i} = (1/2)(1+c_i)\mathbf{h}_{Ei}'\mathbf{h}_{Ei}'^H+\mathbf{I}_{n_E},
\end{equation}
\begin{equation}\label{}
  \mathbf{R}_{i,j} =(1/2)\epsilon_{i,j}\mathbf{h}_{Ei}'\mathbf{h}_{Ej}'^H,
\end{equation}
\begin{equation}\label{}
  \mathbf{R}_{i,M+1} = \sqrt{1/2}\mathbf{h}_{Ei}'\mathbf{r}_{x,i}^H\mathbf{H}_{EA}'^H,
\end{equation}
\begin{equation}\label{}
  \mathbf{R}_{M+1,M+1}=\mathbf{H}_{EA}'\mathbf{H}_{EA}'^H+\mathbf{I}_{n_E},
\end{equation}
where $\epsilon_{i,j}=\mathbb{E}\{\hat p_i\hat p_j^*\}=c_ic_j\phi_{i,j}$ and $\mathbf{r}_{x,i}=\mathbb{E}\{\mathbf{x}\hat p_i^*\}=c_i \mathbf{\bar h}_i^*$.

To obtain an insight into $\sigma_{\Delta s_{i,E}}^2$, let us next choose $i=1$ without loss of generality.
We can rewrite \eqref{eq:R} as
\begin{equation}\label{}
  \mathbf{R}=\left [\begin{array}{cc}
                      \mathbf{R}_{1,1} & \mathbf{\bar R}_1 \\
                      \mathbf{\bar R}_1^H & \mathbf{\bar R}_{1,1}
                    \end{array}
  \right ]
\end{equation}
where $\mathbf{R}_{1,1}$ is the same $n_E\times n_E$ upper-left block of $\mathbf{R}$ in \eqref{eq:R}.
Then
\begin{equation}\label{}
  \mathbf{R}^{-1}=\left [\begin{array}{cc}
                      (\mathbf{R}_{1,1}-\mathbf{\bar R}_1\mathbf{\bar R}_{1,1}^{-1}
                       \mathbf{\bar R}_1^H)^{-1}& * \\
                      * & *
                    \end{array}
  \right ]
\end{equation}
where $*$ denotes matrix blocks of no importance.
Hence, \eqref{eq:s_D_s} with $i=1$ becomes
 \begin{equation}\label{eq:s_Ds1E}
   \sigma_{\Delta s_{1,E}}^2 =
   1-(1/2)\mathbf{h}_{E1}'^H(\mathbf{R}_{1,1}-\mathbf{\bar R}_1\mathbf{\bar R}_{1,1}^{-1}
   \mathbf{\bar R}_1^H)^{-1}
   \mathbf{h}_{E1}'.
 \end{equation}

 Recall
 \begin{equation}\label{}
  \mathbf{R}_{1,1} = (1/2)(1+c_1)\mathbf{h}_{E1}'\mathbf{h}_{E1}'^H+\mathbf{I}_{n_E},
\end{equation}
 \begin{equation}\label{}
  \mathbf{\bar R}_1 =\sqrt{1/2}\mathbf{h}_{E1}'\mathbf{c}_1^H,
\end{equation}
with
  $\mathbf{c}_1^H=\left [\sqrt{1/2}\epsilon_{1,2}\mathbf{h}_{E2}'^H,\cdots,
  \sqrt{1/2}\epsilon_{1,M}\mathbf{h}_{EM}'^H,
  \mathbf{r}_{x,1}^H\mathbf{H}_{EA}'^H\right ]$.
Hence
\begin{equation}\label{}
  \mathbf{\bar R}_1\mathbf{\bar R}_{1,1}^{-1}
                       \mathbf{\bar R}_1^H
                       =(1/2)\mathbf{h}_{E1}'\mathbf{c}_1^H
                       \mathbf{\bar R}_{1,1}^{-1}\mathbf{c}_1\mathbf{h}_{E1}'^H.
\end{equation}

Let
\begin{equation}\label{}
  \gamma_1=1+c_1
  -\mathbf{c}_1^H\mathbf{\bar R}_{1,1}^{-1}\mathbf{c}_1.
\end{equation}
We see that $\gamma_1-1=c_1
  -\mathbf{c}_1^H\mathbf{\bar R}_{1,1}^{-1}\mathbf{c}_1>0$ which is effectively the MSE of the MMSE estimate of $\hat p_1$ by Eve using $\mathbf{y}_{E|1}\doteq[\mathbf{y}_{E,2}^T, \cdots, \mathbf{y}_{E,M}^T, \mathbf{y}_{EA}^T]^T$.

It follows from \eqref{eq:s_Ds1E} that
\begin{align}\label{}
   &\sigma_{\Delta s_{1,E}}^2 =
   1-(1/2)\mathbf{h}_{E1}'^H\left ((1/2)\gamma_1\mathbf{h}_{E1}'\mathbf{h}_{E1}'^H+\mathbf{I}_{n_E}\right)^{-1}
   \mathbf{h}_{E1}'\notag\\
   &=1-\left (\gamma_1\|\mathbf{h}_{E1}'\|^2/2+1\right )^{-1}\|\mathbf{h}_{E1}'\|^2/2\notag\\
   %&=\frac{(\gamma_1-1)
%   \|\mathbf{h}_{E1}'\|^2/2+1}{\gamma_1\|\mathbf{h}_{E1}'\|^2/2+1}\notag\\
   &=\frac{(\gamma_1-1)S_{E,1}/2+1}{\gamma_1S_{E,1}/2+1},
 \end{align}
 with $S_{E,1}=\|\mathbf{h}_{E1}'\|^2$.

Finally, the capacity of the effective return channel from an arbitrary UE, labeled as UE$_1$, to AP relative to $s_1$ is
\begin{align}\label{eq:CE1}
  &C_{E|1} \doteq I(s_1;\mathbf{y}_E)= I(s_1;\hat s_{1,E})= \log(1/\sigma_{\Delta s_{1,E}}^2)\notag\\
   &=\log\left (1+\frac{S_{E,1}/2}{(\gamma_1-1)S_{E,1}/2+1}\right ).
\end{align}

\subsection{Achievable secrecy rate from each UE to AP}
\begin{Proposition}\label{proposition_M-STEEP}
   An achievable secrecy rate of M-STEEP from an arbitrarily selected UE$_1$ to the AP relative to the message symbol $s_1$ from UE$_1$ is $R_{s,1}=(\tilde R_{s,1})^+$ with
  \begin{align}\label{eq:Cs1}
   & \tilde R_{s,1}\doteq I(s_1;\mathbf{y}_A|\mathbf{x})-I(s_1;\mathbf{y}_E)
   \geq  I(s_1;\mathbf{y}_{A1}|\mathbf{x})-I(s_1;\mathbf{y}_E)\notag\\
   &=R_{A|1}-C_{E|1}=\log\left (1+\frac{\frac{S_{A,1}}{2}}{\frac{S_1S_{A,1}/2}{(S_1+1)^2}+1}\right )\notag\\
   &\,\,-\log\left (1+\frac{S_{E,1}/2}{(\gamma_1-1)S_{E,1}/2+1}\right ),
  \end{align}
    where only $\gamma_1$ is affected by UE$_i$'s power for all $i$, i.e., only $\gamma_1$ depends on $S_{E,i}=\|\mathbf{h}_{Ei}'\|^2$ for all $i=1,\cdots,M$.
\end{Proposition}

This proposition follows from \eqref{eq:RAi} and \eqref{eq:CE1}. If $M=1$, it reduces to Corollary \ref{Corollary_1}. 

\begin{Proposition}\label{proposition_gamma1}
 Assume $n_A=1$, rewrite $\mathbf{H}_{EA}'$ as $\mathbf{h}_{EA}'$, and let $S_i=\|\mathbf{h}_i'\|^2$, $S_{Ai}=\|\mathbf{h}_{Ai}'\|^2$ and $S_{Ei}=\|\mathbf{h}_{Ei}'\|^2$, $S_{EA}=\|\mathbf{h}_{EA}'\|^2$, $\alpha_i=\frac{S_{EA}}{S_i}$ and $\beta_i=\frac{S_{Ei}}{S_{Ai}}$.
 Then $\gamma_1-1$ in \eqref{eq:Cs1} becomes
  \begin{align}\label{eq:cond_gamma1}
    &\gamma_1-1=\frac{S_1}{(S_1+1)^2}\left (1+\frac{S_1}{\alpha_1 S_1+1}\left (1-\frac{t_{1,M}}{\alpha_1S_1+1}\right )\right )
  \end{align}
  Also $t_{1,M}=0$ for $M=1$, and $t_{1,M}$ for $M\geq 2$ is defined in \eqref{eq:t_1M} which is a function of $S_{E,A}$ and $S_{E,i}$ for all $i\neq 1$. And $t_{1,M}<\min(M-1,\alpha_1S_1+1)$. Consequently, $R_{s,1}>0$ if and only if
  \begin{equation}\label{eq:cond_2}
   S_{A,1}/2>\left ( 1-\frac{1}{\beta_1}\right )
    \frac{(S_1+1)^2(\alpha_1 S_1+1)}{S_1^2\left (1-\frac{t_{1,M}}{\alpha_1S_1+1}\right )}
 .
\end{equation}
Note that the left side of \eqref{eq:cond_2} is proportional to $p_{u1}$ (the power from UE$_1$) and the right side of \eqref{eq:cond_2} is invariant to $p_{u1}$ and large $p_{ui}$ for all $i\neq 1$.
\end{Proposition}
\begin{IEEEproof}
See Appendix\ref{proof_proposition_gamma1}.
\end{IEEEproof}

This proposition has also been validated by computer simulations.
If $M=1$, \eqref{eq:cond_2} reduces to \eqref{eq:cond1}. But more importantly, we see from \eqref{eq:cond_2} that for any given $M$, the secrecy rate from any UE to AP stays positive if that UE uses a sufficiently large power according to \eqref{eq:cond_2}. Specifically, $1-\frac{t_{1,M}}{\alpha_1S_1+1}$ in \eqref{eq:cond_2} is virtually invariant to a moderate $M$ (e.g., in order of tens) when $\alpha_1S_1$ is large (e.g., in order of 30dB).

\subsection{Total (or sum) secrecy rate of M-STEEP}
\subsubsection{A general expression}
Now define $\mathbf{s}=[s_1,\cdots,s_M]^T$ and recall $\mathbf{y}_A=[\mathbf{y}_{A1}^T,\cdots,\mathbf{y}_{AM}^T]^T$ and $\mathbf{y}_E=[\mathbf{y}_{E1}^T,\cdots,\mathbf{y}_{EM}^T,\mathbf{y}_{EA}^T]^T$. A total achievable secrecy rate of M-STEEP from all UEs to AP (in bits per $M+1$ complex channel uses) can be written as $\tilde R_s^+$ with
\begin{equation}\label{eq:total_secrecy}
  \tilde R_s=I(\mathbf{s};\mathbf{y}_A|\mathbf{x})-I(\mathbf{s};\mathbf{y}_E),
\end{equation}
where the condition on $\mathbf{x}$ in the first term is because of $\mathbf{x}$ being known to AP. This expression \eqref{eq:total_secrecy} is a straightforward extension of the theory behind \eqref{eq:classic_WTC} or \eqref{eq:classic_WTC_2}. In this case, $\mathbf{K}_s=\mathbb{E}\{\mathbf{s}\mathbf{s}^H\}=\mathbf{I}$, and the secret message from each UE is chosen independently (or locally) while the WTC coding scheme can be centrally designed and publicly shared.

Since $I(\mathbf{s};\mathbf{y}_A|\mathbf{x})=\sum_{i=1}^M \tilde R_{s,i|A}$ with $\tilde R_{s,i|A}=I(s_i;\mathbf{y}_A|s_1,\cdots,s_{i-1},\mathbf{x})$, and $I(\mathbf{s};\mathbf{y}_E)=\sum_{i=1}^M \tilde R_{s,i|E}$ with $\tilde R_{s,i|E}=I(s_i;\mathbf{y}_E|s_1,\cdots,s_{i-1})$, we can write
\begin{equation}\label{eq:total_secrecy_1}
  \tilde R_s=\sum_{i=1}^M \tilde R_{s,i}
\end{equation}
with
$\tilde R_{s,i}=\tilde R_{s,i|A}-\tilde R_{s,i|E}$.
Here $\tilde R_{s,i}$ could be negative and its value in general depends on the ordering of the UEs. But $\tilde R_s$ is invariant to the ordering. Also note that $\tilde R_{s,1}$ here is the same as \eqref{eq:Cs1}.

Furthermore, we know
\begin{align}\label{}
  &\tilde R_{s,i|A}\geq
I(s_i;\mathbf{y}_{Ai}|s_1,\cdots,s_{i-1},\mathbf{x})\notag\\
&=I(s_i;\mathbf{y}_{Ai}|\mathbf{x}) =R_{A|i},
\end{align}
which is given by \eqref{eq:RAi}. And
\begin{equation}\label{}
  \tilde R_{s,i|E}
  =-\log\sigma_{\Delta s_{i,E|1:i-1}}^2
\end{equation}
where $\sigma_{\Delta s_{i,E|1:i-1}}^2$ is the MSE of the MMSE estimate of $s_i$ by Eve using $\mathbf{y}_E$ \emph{and} $s_1,\cdots,s_{i-1}$ (as if they are known to Eve), i.e.,
 \begin{equation}\label{eq:s_D_s2}
   \sigma_{\Delta s_{i,E|1:i-1}}^2 =
   1-\mathbf{r}_i^H\mathbf{R}_i^{-1}\mathbf{r}_i
 \end{equation}
 where $\mathbf{r}_i^H =\mathbb{E}\{s_i\mathbf{y}_{E|1:i-1}^H\}$ and $\mathbf{R}_i=\mathbb{E}\{\mathbf{y}_{E|1:i-1}\mathbf{y}_{E|1:i-1}^H\}$.
 Here
 \begin{align}
 &\mathbf{y}_{E|1:i-1}\doteq\notag\\
 &\,\,[\mathbf{\bar y}_{E,1}^T,
 \cdots, \mathbf{\bar y}_{E,i-1}^T,
 \mathbf{y}_{E,i}^T,
 \cdots,\mathbf{y}_{E,M}^T,\mathbf{y}_{E,A}^T]^T
 \end{align}
 with $\mathbf{\bar y}_{E,l}=\mathbf{y}_{E,l}-\mathbf{h}_{El}'s_l$ and $l=1,\cdots,i-1$.
Furthermore, $\mathbf{r}_i$ is given by \eqref{eq:ri}, and
  $\mathbf{R}_i$ is the same as $\mathbf{R}$ in \eqref{eq:R} except that the $l$th diagonal block $\mathbf{R}_{l,l}$ of $\mathbf{R}$ for $l=1,\cdots,i-1$ should be replaced by
$
  \mathbf{\bar R}_{l,l} = \frac{1}{2}c_l\mathbf{h}_{El}'\mathbf{h}_{El}'^H+\mathbf{I}_{n_E}
$.

Next we show a case where all terms in \eqref{eq:total_secrecy_1} can be made positive by a power control even when Eve's receive channels from all other nodes are stronger than those among AP and UEs.

\subsubsection{A special case}
To gain further insights into $\tilde R_s$, let us assume $n_A=n_B=n_E=1$. All channel gains can be now lumped into noise variances. Specifically, in phase 1 of M-STEEP, the AP broadcasts the probe $p$, UE$_i$ receives $y_i=p+w_i$ for $i=1,\cdots,M$, and Eve receives $y_{E,A}=p+w_{E,A}$. Let $\hat p_i$ be the MMSE estimate of $p$ by UE$_i$ from $y_i$. In phase 2 of M-STEEP, UE$_i$ sends $x_i=\hat p_i+s_i$ to the AP using the $i$th orthogonal channel, and hence the AP receives $y_{A,i}=x_i+w_{A,i}$ and Eve receives $y_{E,i}=x_i+w_{E,i}$ for all $i$. Assume $p$ and $s_i$ for all $i$ are i.i.d. $\mathcal{CN}(0,1)$ and all noises ($w_i$, $w_{A,i}$, $w_{E,A}$ and $w_{E,i}$) are i.i.d. $\mathcal{CN}(0,\sigma_*^2)$ with $*$ chosen according to the index of the noise.

It follows that $\hat p_i=(1-\mu_i)y_i$ with $\mu_i=\frac{\sigma_i^2}{1+\sigma_i^2}$, and the MSE of the MMSE estimate of $s_i$ by the AP from $y_{A,i}$ and $p$ is
\begin{equation}\label{eq:AP_M}
  \sigma_{\Delta s_i}^2 = \frac{\mu_i(1-\mu_i)+\sigma_{A,i}^2}{1+\mu_i(1-\mu_i)+\sigma_{A,i}^2}.
\end{equation}
Hence $\tilde R_{s,i|A}\geq -\log\sigma_{\Delta s_i}^2$, the right-side of which is a special case of $R_{A|i}$ in \eqref{eq:RAi} with $p_{ui}=\frac{2}{\sigma_{A,i}^2}$ and $\|\mathbf{h}_{Ai}\|=1$.

One can also verify that the MSE of the MMSE estimate of $s_i$ by Eve from $\mathbf{y}_E\doteq[y_{E,A},y_{E,1},\cdots,y_{E,M}]^T$ is
\begin{equation}\label{}
  \sigma_{\Delta s_{i|E}}^2=1-\mathbf{e}_i^T\left (\mathbf{A}-(1/b)\mathbf{c}\mathbf{c}^T\right )^{-1}\mathbf{e}_i
\end{equation}
where $\mathbf{e}_i$ consists of all zeros except for its $i$th element equal to one, $\mathbf{c}^T=[\mu_1',\cdots,\mu_M']$, $\mu_i'=1-\mu_i$, $b=1+\sigma_{E,A}^2$, and
\begin{equation}\label{}
  \mathbf{A}=\left [\begin{array}{cccc}
                      a_{E,1} & \mu_1'\mu_2' & \cdots & \mu_1'\mu_M' \\
                      \mu_2'\mu_1' & a_{E,2} & \cdots & \cdots \\
                      \cdots & \cdots & \ddots & \mu_{M-1}'\mu_M' \\
                      \mu_M'\mu_1' & \cdots & \mu_M'\mu_{M-1}' & a_{E,M}
                    \end{array}
   \right ]
\end{equation}
with $a_{E,i}=1+\mu_i'+\sigma_{E,i}^2$. Furthermore, the MSE of the MMSE estimate of $s_i$ by Eve from $\mathbf{y}_E$ and $s_1,\cdots,s_{i-1}$ is
\begin{equation}\label{eq:EEE}
  \sigma_{\Delta s_{i|E,1:i-1}}^2
  =1-\mathbf{e}_i^T(\mathbf{A}_i-(1/b)\mathbf{c}\mathbf{c}^T)^{-1}\mathbf{e}_i
\end{equation}
where $\mathbf{A}_i$ is the same as $\mathbf{A}$ after its $l$th diagonal element being replaced by $\mu_l'+\sigma_{E,l}^2$ for $l=1,\cdots,i-1$. Hence
$\tilde R_{s,i|E}=-\log \sigma_{\Delta s_{i|E,1:i-1}}^2$.

\begin{Proposition}\label{Proposition_symmetric}
   For a symmetric network where $\sigma_i=\sigma$, $\sigma_{A,i}=\sigma_A$ and $\sigma_{E,i}=\sigma_E$ for all $i=1,\cdots,M$, the total achievable secrecy rate in \eqref{eq:total_secrecy} is
   \begin{align}\label{eq:total_secrecy_2}
  &\tilde R_s =\tilde R_{s,1}+ \tilde R_{s,2}+\cdots +\tilde R_{s,M}\notag\\
  &\geq \log\frac{\sigma_{\Delta s_{1|E}}^2}{\sigma_{\Delta s_1}^2}
  +\log\frac{\sigma_{\Delta s_{2|E,1}}^2}{\sigma_{\Delta s_2}^2}
  +\cdots+
  \log\frac{\sigma_{\Delta s_{M|E,1:M-1}}^2}{\sigma_{\Delta s_M}^2}.
\end{align}
Referring to the $i$th term after ``$\geq$'' in \eqref{eq:total_secrecy_2} as $R_{s,i}'$, we have $\tilde R_{s,i}\geq R_{s,i}'$ for all $i$, and
\begin{equation}\label{eq:total_secrecy_3}
  R_{s,1}'> R_{s,2}'>\cdots >R_{s,M}'.
\end{equation}
If $\beta_0\doteq\frac{\sigma_A^2}{\sigma_E^2}\leq 1$ (i.e., Eve's channels from UEs are not stronger than AP's channels from UEs), $R_{s,M}'>0$. If $\beta_0$ is fixed and larger than one (i.e., the opposite case from the above and virtually regardless of the channels in phase 1), and $\sigma_A^2$ is inversely proportional to the power $p_B$ used by each UE, then there is a finite threshold $\bar p_B$ such that $R_{s,M}'>0$ if $p_B>\bar p_B$. For $\beta_0>1$ and a large $M$,  $\bar p_B$ increases linearly with $M$.
\end{Proposition}
\begin{IEEEproof}
See Appendix \ref{Proof_Proposition_Symmetric}.
\end{IEEEproof}

Note that $\tilde R_s$ in \eqref{eq:total_secrecy_1} corresponds to a scaled form of all-user (``individual or collective'') secrecy in \cite{TekinYener2008}, $\tilde R_{s,M}$ in \eqref{eq:total_secrecy_1} a scaled form of single-user ``individual secrecy'' in \cite{TekinYener2008}, and  $\tilde R_{s,1}$ in \eqref{eq:Cs1} a scaled form of single-user ``collective secrecy'' in \cite{TekinYener2008}.  While the scheme considered in \cite{TekinYener2008} and \cite{TekinYener2008b} subject to orthogonal access is not able to make $\tilde R_s>0$  when Eve's channels are somewhat stronger than users,  M-STEEP can make all components of $\tilde R_s$ (even the smallest $\tilde R_{s,M}$) positive by choosing strong enough powers from users in phase 2. However, a more complete study of the secrecy rate region subject to power constraints on both AP and users remains open. The assumptions in \cite{TekinYener2008} and \cite{TekinYener2008b} do not cover such a round-trip collaboration exploited by STEEP.

\section{Additional Comments}\label{sec:STEEP_Prior_Ideas}
An important connection between STEEP and the works in \cite{Maurer1993}, \cite{Bloch2011}, \cite{Gamal2011} and \cite{Hayashi2020} has been stressed in the introduction. The idea of feedback for WTC shown in \cite{Lai2008} however has a limited applicability unlike STEEP. The secure feedback channel required in \cite{Bai2022} and many other prior works is not required by STEEP.

The channel probing idea used for phase 1 of STEEP was initially inspired from \cite{HessamMahdavifar2020} and \cite{Li2022} where the authors attempted to use random channel probing to increase the secret-key rate from static reciprocal channels (but with no proven success). This idea is indeed only a special case of the notion of generating correlated data sets at Alice and Bob for SKG \cite{Maurer1993}. The correlated data sets can be directly generated from a channel model as  shown in Chapter 4 in \cite{Bloch2011}. However, the treatment in \cite{Bloch2011} is mostly on the validity and meaning of secret-key capacity and its bounds. It does not address the application of these bounds. The  works shown in \cite{Hayashi2020b} and \cite{Hayashi2022} however addressed SKG in  more applied settings using data sets collected from some specific channel models as well as public communications.

 To achieve the achievable secrecy rate of STEEP, WTC coding is needed in the echoing phase (phase 2). Such coding schemes are available, including LDPC codes \cite{Thangaraj2007} and polar codes \cite{Hessam2011}. However, the current practicality of those codes due to their complexity seems questionable.

 But the effective WTC system constructed by STEEP is such that the user's effective channel is almost surely stronger than Eve's effective channel. Because of this, a positive secrecy rate is virtually given  without the need to know Eve's CSI. Furthermore, to realize a positive secrecy rate for STEEP, we do not necessarily need to use a capacity-achieving channel coding scheme. All we need is a channel code for which the optimal decoding can be done by the receiving user in phase 2. For example,  a convolution code can be used by Bob in phase 2  to encode the stream of the secret information (e.g., $\mathbf{s}$ in \eqref{eq:yA} or $\phi$ in \eqref{eq:yA2}). Then Alice can perform the maximum likelihood decoding, such as Viterbi decoding, of the secret information (e.g., from $\mathbf{\hat s}$ in  \eqref{eq:s_hat_A} or $r_A$ in \eqref{eq:rA}). Since the decoding at Alice is optimal and the effective channel from Bob to Alice is stronger than the effective  channel from Bob to Eve, the error rate at Eve is always higher than that at Alice. The lack of capacity achieving of a channel code would reduce the net channel capacities for both user and Eve but without necessarily a significant change to a positive secrecy rate. For a Gaussian-noise channel, the error rate drops exponentially as SNR increases, which creates a drastic difference between the number of errors at Alice and those at Eve. Such a gap of error rates can be used as a secrecy measure.

 Provided no error is detected at Alice (using any of the established channel codes), if the secret information is meant to generate a secret key, a hash function could then be applied at Alice and Bob to produce the secret key with a higher confidence of its secrecy (also known as privacy amplification). The secret-key rate in bits/s/Hz of this STEEP-assisted method for SKG does not reduce to zero as the channel coherence time increases, unlike numerous methods in the literature such as \cite{Wilson2007} and \cite{Wallace2010} based on reciprocal channels. To know the exact amount of secrecy, it would always require the knowledge of Eve's channel. But it could suffice in practice  that there is at least some amount of positive secrecy rate even in the worst possible case.

 STEEP may also remind one of a widely used method for networking security called ``nonce''.  The usefulness of nonce is based on the assumption that Alice can send a nonce reliably to Bob while Eve can not receive it. Then this nonce can be used (normally once) by Bob to encrypt a message to be sent to Alice. Unlike nonce, STEEP allows Eve to receive the probes from Alice but with some noise while Bob does not have to receive the probes with more accuracy than Eve, and the noisy probes received by Bob are used to encrypt a secret message to be sent to Alice. STEEP is naturally applicable at the physical layer due to presence of independent noises (especially thermal noises)
 while its applicability at a higher layer is also of great interest.

 \section{Conclusion}
 Although related to some contributions in \cite{Maurer1993}, \cite{Bloch2011}, \cite{Gamal2011} and \cite{Hayashi2020}, STEEP as shown in this paper has a broad applicability for secure communications at the physical layer. This paper has presented: STEEP based on Gaussian probing signals and Gaussian linear encryption over Gaussian MIMO channels (G-STEEP); STEEP based on phase-shift-keying (PSK) probing signals and nonlinear PSK encryption over Gaussian SISO channel (P-STEEP); and a special form of G-STEEP for orthogonal multiple access between an access point and multiple users (M-STEEP). Achievable secrecy rates of these schemes have been derived and analyzed. It has been shown that positive secrecy rates for both single-link problem and multiple-access problem can be virtually guaranteed by asymmetric power allocation as long as Eve's receive channel in the probing phase of STEEP is not noiseless, which includes the secrecy rate from a user to AP subject to exposure of messages from all other users. Such a discovery is highly novel, and in great contrast to numerous works in the physical layer security literature over past three decades that require user's (including AP here and below) receive channel being stronger than Eve's, user's  antennas more than Eve's, reciprocal channel responses between users, secure feedback channel between users, collaborative third party (such as relay), and/or in-band full-duplex between users, in order to ensure a positive secrecy rate between users. While rooted in  the encryption lemma discussed in the introduction, STEEP that exploits echoing encrypted probes, asymmetric power allocation and/or collaborative round-trip coding should have opened a new door of research and development for secure communications.

%\section*{Acknowledgement}
%The author appreciates many comments from anonymous reviewers.

\section*{Appendix}
\begin{appendices}

\subsection{Proof of proposition \ref{Proposition_DoF}}\label{sec:Proof_Proposition_DoF}
We will assume $n_A\geq n_B$.
Recall $\mathbf{R}_{\Delta \mathbf{p}'}$ in \eqref{eq:RDpP}, $\mathbf{R}_{\Delta \mathbf{p}}$ in \eqref{eq:R_Delta_p} and $C_{A|B,G}=\log\frac{N_{A|B}}{D_{A|B}}$ in \eqref{eq:CAB_MIMO}. Also recall $\mathbf{H}_{AB}''=\sqrt{p_B/(2n_B)}\mathbf{H}_{AB}$, $\mathbf{H}_{EB}''=\sqrt{p_B/(2n_B)}\mathbf{H}_{EB}$, $\mathbf{H}_{BA}'=\sqrt{p_A/n_A}\mathbf{H}_{BA}$, and $\mathbf{H}_{EA}'=\sqrt{p_A/n_A}\mathbf{H}_{EA}$.

We know $\lim_{p_A\to\infty}\mathbf{R}_{\Delta \mathbf{p}'}=\lim_{p_A\to\infty}\mathbf{R}_{\Delta \mathbf{p}}= \mathbf{0}$. Then, subject to a fixed $\eta_p=\frac{p_B}{p_A}$,
\begin{align}\label{}
  &\lim_{p_A\to\infty} \log D_{A|B}=\log\left |\frac{\eta_pn_A}{2n_B}\mathbf{H}_{AB}^H\mathbf{H}_{AB}
(\boldsymbol{\Pi}_{BA}^2)^{-1}+
\mathbf{I}_{n_B}\right |
\end{align}
 which is invariant to $p_A$, and
 \begin{align}
 &\lim_{p_A\to\infty}\log N_{A|B}=\lim_{p_A\to\infty}\log\left |\frac{p_B}{2n_B}\mathbf{H}_{AB}^H\mathbf{H}_{AB}\right |+o(\log p_A).
 \end{align}
  Here $o(x)$ is a quantity such that $\lim_{x\to\infty}\frac{o(x)}{x}=0$. And therefore, for $n_A\geq n_B$ and a fixed $\eta_p=\frac{p_B}{p_A}$,
  \begin{equation}\label{}
    \lim_{p_A\to\infty}\log N_{A|B}/\log p_A=n_B,
  \end{equation}
  \begin{equation}\label{}
    \lim_{p_A\to\infty}\log D_{A|B}/\log p_A=0,
  \end{equation}
  and hence
\begin{equation}\label{eq:DoF_CAB}
  \lim_{p_A\to\infty}C_{A|B,G}/\log p_A=n_B.
\end{equation}

Now let us consider $C_{E|B,G}=\log\frac{N_{E|B}}{D_{E|B}}$ in \eqref{eq:CEB_MIMO}  and $\mathbf{T}$ in \eqref{eq:T}. We know that $rank(\mathbf{H}_{EA})=\min(n_A,n_E)\doteq r_A$. We can write the eigenvalue decomposition (EVD) of $\mathbf{H}_{EA}^H\mathbf{H}_{EA}$ as $\mathbf{U}\mathbf{D}^2\mathbf{U}^H=
\mathbf{U}_A\mathbf{D}_A^2\mathbf{U}_A^H$ where $\mathbf{D}_A^2$ is $r_A\times r_A$ nonsingular diagonal and $\mathbf{U}_A$ is the corresponding $r_A$ columns of the $n_A\times n_A$ unitary matrix $\mathbf{U}$. It follows from \eqref{eq:T} that
\begin{align}\label{}
  &\mathbf{T} = \frac{p_A}{n_A}\mathbf{R}_{\mathbf{\hat p}}^H\mathbf{V}_{BA}^H\mathbf{U}
  \left (\frac{p_A}{n_A}\mathbf{D}^2+\mathbf{I}_{n_A}\right )^{-1}\notag\\
  &\,\,\cdot\mathbf{D}^2\mathbf{U}^H\mathbf{V}_{BA}\mathbf{R}_{\mathbf{\hat p}}\notag\\
  &=\frac{p_A}{n_A}\mathbf{R}_{\mathbf{\hat p}}^H\mathbf{V}_{BA}^H\mathbf{U}_A
  \left (\frac{p_A}{n_A}\mathbf{D}_A^2+\mathbf{I}_{r_A}\right )^{-1}\notag\\
  &\,\,\cdot\mathbf{D}_A^2\mathbf{U}_A^H\mathbf{V}_{BA}\mathbf{R}_{\mathbf{\hat p}}.
\end{align}
Hence,
\begin{equation}\label{}
  \lim_{p_A\to\infty}\mathbf{T}
  =\mathbf{V}_{BA}^H\mathbf{U}_A
  \mathbf{U}_A^H\mathbf{V}_{BA}
\end{equation}
where we have used $\lim_{p_A\to\infty}\mathbf{R}_{\mathbf{\hat p}}=\mathbf{I}_{n_B}$.

Since $\mathbf{R}_{\Delta \mathbf{\hat p}_E}=\mathbf{R}_{\mathbf{\hat p}}-\mathbf{T}$ as in \eqref{eq:R_Delta_hat_pE}, then for any $p_B$,
\begin{align}\label{}
  &\lim_{p_A\to\infty}\mathbf{R}_{\Delta \mathbf{\hat p}_E}=\mathbf{I}_{n_B}-
  \mathbf{V}_{BA}^H\mathbf{U}_A
  \mathbf{U}_A^H\mathbf{V}_{BA}\notag\\
 & =\mathbf{V}_{BA}^H\mathbf{P}_A^\perp
  \mathbf{V}_{BA},
\end{align}
where $\mathbf{P}_A^\perp=\mathbf{I}_{n_A}-\mathbf{U}_A
  \mathbf{U}_A^H$ is the projection matrix onto the orthogonal complement of $range(\mathbf{U}_A)$, and has the rank $(n_A-n_E)^+$. Furthermore,
\begin{align}\label{}
  &\lim_{p_A\to\infty}\log D_{E|B}\notag\\
  &
  =\log\left |\frac{p_B}{2n_B}\mathbf{H}_{EB}^H\mathbf{H}_{EB}\mathbf{V}_{BA}^H\mathbf{P}_A^\perp\mathbf{V}_{BA}
  +\mathbf{I}_{n_B}\right |,
\end{align}
\begin{align}\label{}
  &\lim_{p_A\to\infty}\log N_{E|B}
  \notag\\
  &=\log\left |\frac{p_B}{2n_B}\mathbf{H}_{EB}^H\mathbf{H}_{EB}
  (\mathbf{V}_{BA}^H\mathbf{P}_A^\perp\mathbf{V}_{BA}+\mathbf{I}_{n_B})
  +\mathbf{I}_{n_B}\right |.
\end{align}
It is typical (or with probability one for  random matrices) that
\begin{align}\label{eq:rank_cond_1}
&rank(\mathbf{H}_{EB}^H\mathbf{H}_{EB}\mathbf{V}_{BA}^H\mathbf{P}_A^\perp\mathbf{V}_{BA})\notag\\
&=\min(rank(\mathbf{H}_{EB}),rank(\mathbf{V}_{BA}),rank(\mathbf{P}_A^\perp))\notag\\
&=\min(n_B,(n_A-n_E)^+),
\end{align}
and
\begin{equation}\label{eq:rank_cond_2}
  rank(\mathbf{H}_{EB}^H\mathbf{H}_{EB}\mathbf{V}_{BA}^H\mathbf{P}_A^\perp\mathbf{V}_{BA}+\mathbf{I}_{n_B})
=n_B.
\end{equation}
Therefore,  for fixed $\eta_p=\frac{p_B}{p_A}$,
\begin{equation}\label{}
  \lim_{p_A\to\infty}N_{E|B}/\log p_A=n_B,
\end{equation}
\begin{equation}\label{}
  \lim_{p_A\to\infty}D_{E|B}/\log p_A=\min(n_B,(n_A-n_E)^+),
\end{equation}
and hence
\begin{equation}\label{eq:DoF_CEB}
  \lim_{p_A\to\infty}\frac{C_{E|B,G}}{\log p_A} = n_B-\min(n_B,(n_A-n_E)^+).
\end{equation}
Applying  \eqref{eq:DoF_CAB} and \eqref{eq:DoF_CEB} yields that for fixed $\eta_p=\frac{p_B}{p_A}$,
\begin{equation}\label{}
  \lim_{p_A\to\infty}\frac{1}{\log p_A}R_{s,G} = \min(n_B,(n_A-n_E)^+).
\end{equation}

Finally, we rewrite \eqref{eq:MIMO_key} as
\begin{align}\label{eq:MIMO_key_2}
  &C_{key}=\log\left |\mathbf{I}_{n_A}+\frac{p_A}{n_A}\mathbf{\tilde H}_{EA}^H\mathbf{\tilde H}_{EA}
  \right |
  -\log\left |\frac{p_A}{n_A}\mathbf{H}_{EA}^H\mathbf{H}_{EA}+\mathbf{I}_{n_A}\right |
\end{align}
with $\mathbf{\tilde H}_{EA}=\left [ \begin{array}{c}
                                    \mathbf{H}_{EA} \\
                                    \mathbf{H}_{BA}
                                  \end{array}
\right ]$. Here,  $rank(\mathbf{H}_{EA})=\min(n_A,n_E)$ and
\begin{equation}\label{eq:rank_cond_3}
  rank(\mathbf{\tilde H}_{EA})=\min(n_A,n_E+n_B).
\end{equation}
 It follows that
\begin{align}\label{}
  &\lim_{p_A\to\infty}\frac{1}{\log p_A}C_{key}=\min(n_A,n_E+n_B)-\min(n_A,n_E)\notag\\
  &=\min(n_B,(n_A-n_E)^+).
\end{align}
The proof is completed.

\subsection{Proof of proposition \ref{Proposition_Ckey_Cs}}\label{sec:Proof_Proposition_Ckey_Cs}
It is easy to verify that $C_{key}$ from \eqref{eq:MIMO_key} satisfies the second equation in \eqref{eq:MIMO_optimal}. We will next show the first equation in \eqref{eq:MIMO_optimal}.

For $n_E\geq n_A\geq n_B$, both $\mathbf{H}_{AB}^H\mathbf{H}_{AB}$ and $\mathbf{H}_{EB}^H\mathbf{H}_{EB}$ have full rank $n_B$, and hence \eqref{eq:MIMO_Cs_new} implies
\begin{align}\label{}
  &R_{s,G}^B\doteq \lim_{p_B\to\infty}R_{s,G}
  \notag\\
  &=\log\left (\frac{|
   (\mathbf{R}_{\Delta\mathbf{p}'}+\mathbf{I}_{n_B})|}{|
  \mathbf{R}_{\Delta\mathbf{p}'}|}\frac{|
  \mathbf{R}_{\Delta\mathbf{\hat p}_E}
  |}{|
  (\mathbf{R}_{\Delta\mathbf{\hat p}_E}+\mathbf{I}_{n_B})|}\right ).
\end{align}

Next we consider $R_{s,G}^{B,A}\doteq \lim_{p_A\to\infty}R_{s,G}^B$.
Since $\mathbf{H}_{EA}^H\mathbf{H}_{EA}$ is invertible, then $\lim_{p_A\to\infty}\mathbf{T}=\mathbf{I}_{n_B}$ and $\lim_{p_A\to\infty}\mathbf{R}_{\Delta\mathbf{\hat p}_E}=0$. Also note that
$\mathbf{R}_{\mathbf{\hat p}}
  =\mathbf{I}_{n_B}+\mathcal{O}(1/p_A)$ and $\lim_{p_A\to\infty}\mathbf{R}_{\Delta\mathbf{p}'}=\lim_{p_A\to\infty}\mathbf{R}_{\Delta\mathbf{p}}=0$. Therefore,
\begin{align}\label{}
  &R_{s,G}^{B,A}\doteq \lim_{p_A\to\infty}R_{s,G}^B=\lim_{p_A\to\infty}\log\left (\frac{|
  \mathbf{R}_{\Delta\mathbf{\hat p}_E}
  |}{|
  \mathbf{R}_{\Delta\mathbf{p}}|}\right )
\end{align}
where $\mathbf{R}_{\Delta\mathbf{p}'}=\mathbf{R}_{\Delta\mathbf{p}}(\mathbf{I}+\mathcal{O}(1/p_A))$ for large $p_A$ has been used, and the indefinite form of $\frac{0}{0}$ is resolved next.

Let $\gamma_A\doteq\frac{p_A}{n_A}$, $\mathbf{W}_E\doteq\mathbf{H}_{EA}^H\mathbf{H}_{EA}$ and $\mathbf{T}_P\doteq\gamma_A\boldsymbol{\Pi}_{BA}^2+\mathbf{I}_{n_B}$. Then
\begin{align}
&\mathbf{R}_{\Delta\mathbf{p}}^{-1}\mathbf{R}_{\Delta\mathbf{\hat p}_E}
=\mathbf{T}_P\left (\gamma_A\boldsymbol{\Pi}_{BA}^2
  \mathbf{T}_P^{-1}\right . \notag\\
  &\,\,\left .-\gamma_A\mathbf{R}_{\mathbf{\hat p}}^H\mathbf{V}_{BA}^H\left (\gamma_A\mathbf{W}_E+\mathbf{I}_{n_E}\right )^{-1}
\mathbf{W}_E\mathbf{V}_{BA}\mathbf{R}_{\mathbf{\hat p}}\right )\notag\\
&=\gamma_A\boldsymbol{\Pi}_{BA}^2
   -\gamma_A^2\boldsymbol{\Pi}_{BA}^2\mathbf{V}_{BA}^H\left (\gamma_A\mathbf{W}_E+\mathbf{I}_{n_E}\right )^{-1}\notag\\
  &\,\,\cdot
\mathbf{W}_E\mathbf{V}_{BA}\mathbf{R}_{\mathbf{\hat p}}
\end{align}
Also note
\begin{align}\label{}
  &(\gamma_A\mathbf{W}_E+\mathbf{I}_{n_E})^{-1}
  =\gamma_A^{-1}\mathbf{W}_E^{-1}\notag\\
  &\,\,-\gamma_A^{-1}\mathbf{W}_E^{-1}(\mathbf{I}_{n_E}
  +\gamma_A^{-1}\mathbf{W}_E^{-1})^{-1}\gamma_A^{-1}\mathbf{W}_E^{-1},
\end{align}
\begin{align}\label{}
  &\mathbf{R}_{\mathbf{\hat p}}=\gamma_A\boldsymbol{\Pi}_{BA}^2(\gamma_A\boldsymbol{\Pi}_{BA}^2+\mathbf{I}_{n_B})^{-1}
  \notag\\
  &=\mathbf{I}_{n_B}-(\mathbf{I}_{n_B}+
  \gamma_A^{-1}\boldsymbol{\Pi}_{BA}^{-2})^{-1}\gamma_A^{-1}\boldsymbol{\Pi}_{BA}^{-2}.
\end{align}
Then, one can verify that
\begin{align}
&\mathbf{R}_{\Delta\mathbf{p}}^{-1}\mathbf{R}_{\Delta\mathbf{\hat p}_E}
=(\mathbf{I}_{n_B}+\gamma_A^{-1}\boldsymbol{\Pi}_{BA}^{-2})^{-1}
+\notag\\
&\,\,\boldsymbol{\Pi}_{BA}^2\mathbf{V}_{BA}^H\mathbf{W}_E^{-1}(\mathbf{I}_{n_A}+
\gamma_A^{-1}
\mathbf{W}_E^{-1})^{-1}\mathbf{V}_{BA}\mathbf{R}_{\mathbf{\hat p}},
\end{align}
and therefore
\begin{align}\label{}
  &R_{s,G}^{B,A}=\lim_{p_A\to\infty}\log |\mathbf{R}_{\Delta\mathbf{p}}^{-1}
  \mathbf{R}_{\Delta\mathbf{\hat p}_E}|\notag\\
  &=\log\left |\mathbf{I}_{n_B}+\boldsymbol{\Pi}_{BA}^2\mathbf{V}_{BA}^H
  \mathbf{W}_E^{-1}\mathbf{V}_{BA}\right |
\end{align}
which completes the proof.
\subsection{Proof of proposition \ref{proposition_gamma1}}\label{proof_proposition_gamma1}
It is easy to verify  that subject to \eqref{eq:cond_gamma1}, $R_{s,1}>0$ if and only if
 \begin{equation}\label{eq:cond_1}
    \frac{S_{A,1}}{2}\left (1-\frac{t_{1,M}}{\alpha_1S_1+1}\right )>\left ( 1-\frac{1}{\beta_1}\right )
    \frac{(S_1+1)^2(\alpha_1 S_1+1)}{S_1^2}.
  \end{equation}

It will be obvious that for $M=1$, $t_{1,M}=0$. We will also see that
$t_{1,M}$ is an increasing function of $S_{E,A}$ while
 $\frac{t_{1,M}}{S_{E,A}+1}$ is a decreasing function of $S_{E,A}$.

We will also show explicitly that $t_{1,M}<M-1$. This means that if $S_{E,A}=\alpha_1 S_1>M-2$, then $1-\frac{t_{1,M}}{\alpha_1S_1+1}>0$ and hence \eqref{eq:cond_1} is equivalent to \eqref{eq:cond_2}. Since $ R_{s,1}$ must be a decreasing function of $\alpha_1$ (which is the ratio of the channel strength from AP to Eve over that from AP to UE$_1$), $ R_{s,1}$ must increase as $\alpha_1$ decreases. If the peak value of $\frac{t_{1,M}}{\alpha_1S_1+1}$ (a decreasing function of $\alpha_1$) is larger than one, then as $\alpha_1$ decreases (starting from the condition $S_{E,A}=\alpha_1 S_1>M-2$) the condition $1-\frac{t_{1,M}}{\alpha_1S_1+1}>0$ would be reversed and a contradiction from $ R_{s,1}>0$ would be concluded. Therefore, $t_{1,M}<\alpha_1S_1+1$ must hold for all $\alpha_1$. This also means that $t_{1,M}<1$ when $S_{E,A}=\alpha_1S_1=0$.

We will only need to prove \eqref{eq:cond_gamma1} for all $M$, and $t_{1,M}<M-1$ for $M\geq 2$.
Recall
$\gamma_1-1=c_1-q_1$
with
\begin{equation}\label{eq:q1_1}
  q_1 = \mathbf{c}_1^H\mathbf{\bar R}_{1,1}^{-1}\mathbf{c}_1.
\end{equation}
Also recall $c_i=\sigma_{\hat p_i}^2=\mathbb{E}\{|\hat p_i|^2\}$; and for $n_A=1$, $\phi_{i,j}=1$ and hence $\epsilon_{i,j}=c_ic_j$.

To simplify the notions, we will use $\mathbf{g}_i\doteq\sqrt{1/2}\mathbf{h}_{Ei}'$ and $\mathbf{g}_A\doteq\mathbf{h}_{EA}'$.
 Then,
 \begin{equation}\label{eq:vector_c1}
   \mathbf{c}_1^H=c_1[c_2\mathbf{g}_2^H,c_3\mathbf{g}_3^H, \cdots, c_M\mathbf{g}_M^H|\mathbf{g}_A^H]
   \doteq c_1[\mathbf{c}_a^H|\mathbf{c}_b^H]
 \end{equation}
 with $\mathbf{c}_a=[c_2\mathbf{g}_2^H,c_3\mathbf{g}_3^H, \cdots, c_M\mathbf{g}_M^H]^H$ and $\mathbf{c}_b=\mathbf{g}_A$. Also
 \begin{align}\label{}
   &\mathbf{\bar R}_{1,1}
   =\mathbf{I}+\notag\\
   &\left[
   \begin{array}{ccc|c}
     (1+c_2^2)\mathbf{g}_2\mathbf{g}_2^H  &\cdots&c_2c_M\mathbf{g}_2\mathbf{g}_M^H &c_2\mathbf{g}_2\mathbf{g}_A^H \\
     \cdots&\cdots&\cdots&\cdots\\
     c_Mc_2\mathbf{g}_M\mathbf{g}_2^H  & \cdots&
     (1+c_M^2)\mathbf{g}_M\mathbf{g}_M^H&c_M\mathbf{g}_M\mathbf{g}_A^H   \\
      \hline
      c_2\mathbf{g}_A\mathbf{g}_2^H&  \cdots& c_M\mathbf{g}_A\mathbf{g}_M^H& \mathbf{g}_A\mathbf{g}_A^H
   \end{array}
   \right ]\notag\\
   &
   \doteq\left [ \begin{array}{c|c}
              \mathbf{A}_{1,1} & \mathbf{A}_{1,2} \\
              \hline
              \mathbf{A}_{1,2}^H & \mathbf{A}_{2,2}
            \end{array}
   \right ]
 \end{align}
 with $\mathbf{A}_{1,2}=\mathbf{c}_a\mathbf{g}_A^H$ and $\mathbf{A}_{2,2}=\mathbf{g}_A\mathbf{g}_A^H+\mathbf{I}$.
 Also let
 \begin{equation}\label{eq:bar_R11_inverse}
   \mathbf{\bar R}_{1,1}^{-1}
   =\left [\begin{array}{cc}
             \mathbf{B}_{1,1} & \mathbf{B}_{1,2} \\
             \mathbf{B}_{1,2}^H & \mathbf{B}_{2,2}
           \end{array}
    \right ]
 \end{equation}
 Then it is known that
 \begin{equation}\label{eq:B11}
   \mathbf{B}_{1,1}
   =(\mathbf{A}_{1,1}-\mathbf{A}_{1,2}\mathbf{A}_{2,2}^{-1}\mathbf{A}_{1,2}^H)^{-1},
 \end{equation}
\begin{equation}\label{}
  \mathbf{B}_{2,2}
  =\mathbf{A}_{2,2}^{-1}+\mathbf{A}_{2,2}^{-1}\mathbf{A}_{1,2}^H\mathbf{B}_{1,1}
  \mathbf{A}_{1,2}\mathbf{A}_{2,2}^{-1},
\end{equation}
\begin{equation}\label{}
  \mathbf{B}_{1,2}=-\mathbf{B}_{1,1}\mathbf{A}_{1,2}\mathbf{A}_{2,2}^{-1}.
\end{equation}

Here
\begin{align}
&\mathbf{A}_{1,2}\mathbf{A}_{2,2}^{-1}\mathbf{A}_{1,2}^H
=\mathbf{c}_a\mathbf{g}_A^H(\mathbf{g}_A\mathbf{g}_A^H+\mathbf{I})^{-1}
\mathbf{g}_A\mathbf{c}_a^H\notag\\
&=\frac{S_{E,A}}{S_{E,A}+1}\mathbf{c}_a\mathbf{c}_a^H
\end{align}
with $S_{E,A}=\|\mathbf{g}_A\|^2$. Then, it follows from \eqref{eq:B11} that
\begin{align}
&\mathbf{B}_{1,1}
   =\left (\mathbf{I}+\right .\notag\\
   &\left .\left [\begin{array}{ccc}
                    \left (1+\frac{c_2^2}{S_{E,A}+1}\right )\mathbf{g}_2\mathbf{g}_2^H &\cdots& \frac{c_2c_M}{S_{E,A}+1}\mathbf{g}_2\mathbf{g}_M^H\\
                     \cdots&\cdots&\cdots\\
                     \frac{c_Mc_2}{S_{E,A}+1}\mathbf{g}_M\mathbf{g}_2^H
                     &\cdots&\left (1+\frac{c_M^2}{S_{E,A}+1}\right )\mathbf{g}_M\mathbf{g}_M^H
                  \end{array}
    \right ]\right )^{-1}
\end{align}

We know from \eqref{eq:q1_1}, \eqref{eq:vector_c1} and \eqref{eq:bar_R11_inverse} that
\begin{equation}\label{eq:q1_2}
  q_1 = c_1^2(q_a+2\Re\{q_b\}+q_c)
\end{equation}
with $q_a=\mathbf{c}_a^H\mathbf{B}_{1,1}\mathbf{c}_a$, $q_b=\mathbf{c}_a^H\mathbf{B}_{1,2}\mathbf{c}_b$ and $q_c=
  \mathbf{c}_b^H\mathbf{B}_{2,2}\mathbf{c}_b$.

It follows that
\begin{align}
&q_b
=-\mathbf{c}_a^H\mathbf{B}_{1,1}\mathbf{c}_a\mathbf{g}_A^H(\mathbf{g}_A\mathbf{g}_A^H+
\mathbf{I})^{-1}\mathbf{g}_A=-q_a\frac{S_{E,A}}{S_{E,A}+1}.
\end{align}
And
\begin{align}
&q_c=\mathbf{g}_A^H\left ((\mathbf{g}_A\mathbf{g}_A^H+\mathbf{I})^{-1}+\right .\notag\\
&\,\,\left .
 (\mathbf{g}_A\mathbf{g}_A^H+\mathbf{I})^{-1}\mathbf{g}_A\mathbf{c}_a^H\mathbf{B}_{1,1}
 \mathbf{c}_a\mathbf{g}_A^H(\mathbf{g}_A\mathbf{g}_A^H+\mathbf{I})^{-1}\right )\mathbf{g}_A\notag\\
&=\left (\frac{S_{E,A}}{S_{E,A}+1}+\frac{S_{E,A}^2}{(S_{E,A}+1)^2}
\mathbf{c}_a^H\mathbf{B}_{1,1}\mathbf{c}_a\right )\notag\\
&=\frac{S_{E,A}}{S_{E,A}+1}+\frac{S_{E,A}^2}{(S_{E,A}+1)^2}q_a.
\end{align}

Therefore, \eqref{eq:q1_2} becomes
\begin{align}\label{eq:q1}
&q_1 = c_1^2\left (q_a-2q_a\frac{S_{E,A}}{S_{E,A}+1}+\frac{S_{E,A}}{S_{E,A}+1}+
\frac{S_{E,A}^2}{(S_{E,A}+1)^2}q_a\right )\notag\\
&=c_1^2\left (\frac{q_a}{(S_{E,A}+1)^2}+\frac{S_{E,A}}{S_{E,A}+1}\right ).
\end{align}

We now let $t_{1,M}\doteq q_a$. It is obvious that $t_{1,M}=0$ for $M=1$. We will show $t_{1,M}<M-1$ for $M\geq 2$.
We can also write
\begin{equation}\label{eq:t_1M}
  t_{1,M} = \mathbf{v}_M^H\mathbf{B}_M^{-1} \mathbf{v}_M
\end{equation}
with
\begin{equation}\label{}
  \mathbf{v}_M^H=[c_2\mathbf{g}_2^H,\cdots,c_{M-1}\mathbf{g}_{M-1}^H|c_M\mathbf{g}_M^H]
  =[\mathbf{v}_{M-1}^H|c_M\mathbf{g}_M^H],
\end{equation}
\begin{align}
&\mathbf{B}_M=\mathbf{I}+\notag\\
&\left [\begin{array}{cc|c}
                    \left (1+\frac{c_2^2}{S_{E,A}+1}\right )\mathbf{g}_2\mathbf{g}_2^H & \cdots& \frac{c_2c_M}{S_{E,A}+1}\mathbf{g}_2\mathbf{g}_M^H \\
                    \cdots&\cdots&\cdots\\
                    \hline
                     \frac{c_Mc_2}{S_{E,A}+1}\mathbf{g}_M\mathbf{g}_2^H& \cdots& \left (1+\frac{c_M^2}{S_{E,A}+1}\right )\mathbf{g}_M\mathbf{g}_M^H
                  \end{array}
    \right ]\notag\\
    &=\left [ \begin{array}{c|c}
                \mathbf{B}_{M-1} & \mathbf{C}_{M-1} \\
                \hline
                \mathbf{C}_{M-1}^H & \left (1+\frac{c_M^2}{S_{E,A}+1}\right )\mathbf{g}_M\mathbf{g}_M^H+\mathbf{I}
              \end{array}
    \right ].
\end{align}
Here $\mathbf{C}_{M-1}=\frac{c_M}{S_{E,A}+1}\mathbf{v}_{M-1}\mathbf{g}_M^H$.

We see that for $M=2$,
\begin{align}\label{}
  &t_{1,2} = \mathbf{v}_2^H\mathbf{B}_2^{-1}\mathbf{v}_2 =\frac{c_2^2S_{E,2}}{(1+\frac{c_2^2}{S_{E,A}+1})S_{E,2}+1}\notag\\
  &<
  \frac{c_2^2}{(1+\frac{c_2^2}{S_{E,A}+1})}<c_2^2.
\end{align}

For $M>2$,
\begin{equation}\label{}
  \mathbf{B}_M^{-1}=\left [\begin{array}{cc}
                             \mathbf{B}_{M|1,1} & \mathbf{B}_{M|1,2} \\
                             \mathbf{B}_{M|1,2}^H & \mathbf{B}_{M|2,2}
                           \end{array}
  \right ]
\end{equation}
where
\begin{align}\label{}
  &\mathbf{B}_{M|2,2}=\left (\left (1+\frac{c_M^2}{S_{E,A}+1}\right )\mathbf{g}_M\mathbf{g}_M^H+\mathbf{I}\right .\notag\\
  &\,\,\left .-\mathbf{C}_{M-1}^H\mathbf{B}_{M-1}^{-1}
  \mathbf{C}_{M-1}\right )^{-1},
\end{align}
\begin{equation}\label{}
  \mathbf{B}_{M|1,1}
  =\mathbf{B}_{M-1}^{-1}-\mathbf{B}_{M-1}^{-1}\mathbf{C}_{M-1}
  \mathbf{B}_{M|2,2}\mathbf{C}_{M-1}^H\mathbf{B}_{M-1}^{-1},
\end{equation}
\begin{equation}\label{}
  \mathbf{B}_{M|1,2}=-\mathbf{B}_{M-1}^{-1}\mathbf{C}_{M-1}\mathbf{B}_{M|2,2}.
\end{equation}

Then,
\begin{align}
&t_{1,M} =v_M+2\Re\{w_M\}+\eta_M
\end{align}
with $v_M = \mathbf{v}_{M-1}^H\mathbf{B}_{M|1,1}\mathbf{v}_{M-1}$, $w_M=\mathbf{v}_{M-1}^H\mathbf{B}_{M|1,2}c_M\mathbf{g}_M$, and $\eta_M = c_M^2
\mathbf{g}_M^H\mathbf{B}_{M|2,2}\mathbf{g}_M$.

We know
\begin{align}
&\mathbf{C}_{M-1}^H\mathbf{B}_{M-1}^{-1}
  \mathbf{C}_{M-1}
  =\frac{c_M^2}{(S_{E,A}+1)^2}\mathbf{g}_M\mathbf{v}_{M-1}^H\mathbf{B}_{M-1}^{-1}
  \mathbf{v}_{M-1}\mathbf{g}_M^H\notag\\
  &=\frac{t_{1,M-1}c_M^2}{(S_{E,A}+1)^2}\mathbf{g}_M\mathbf{g}_M^H,
\end{align}
and hence
\begin{equation}\label{}
  \mathbf{B}_{M|2,2}=\left (\left (1+\frac{c_M^2}{S_{E,A}+1}-\frac{t_{1,M-1}c_M^2}{(S_{E,A}+1)^2}\right )\mathbf{g}_M\mathbf{g}_M^H+\mathbf{I}\right )^{-1}.
\end{equation}

Then
\begin{align}\label{eq:eta_M}
  &\eta_M = \frac{c_M^2S_{E,M}}{\left (1+\frac{c_M^2}{S_{E,A}+1}-\frac{t_{1,M-1}c_M^2}{(S_{E,A}+1)^2}\right )S_{E,M}+1}\notag\\
  &<\frac{c_M^2}{\left (1+\frac{c_M^2}{S_{E,A}+1}-\frac{t_{1,M-1}c_M^2}{(S_{E,A}+1)^2}\right )}.
\end{align}
This bound is tight when $S_{E,M}$ is large. Also,
\begin{align}
&v_M = \mathbf{v}_{M-1}^H\left (\mathbf{B}_{M-1}^{-1}\right .\notag\\
&\,\,\left .-\mathbf{B}_{M-1}^{-1}\mathbf{C}_{M-1}
  \mathbf{B}_{M|2,2}\mathbf{C}_{M-1}^H\mathbf{B}_{M-1}^{-1}
  \right )\mathbf{v}_{M-1}\notag\\
  &=t_{1,M-1}-t_{1,M-1}^2\frac{c_M^2}{(S_{E,A}+1)^2}\mathbf{g}_M^H\mathbf{B}_{M|2,2}\mathbf{g}_M\notag\\
  &=t_{1,M-1}-t_{1,M-1}^2\frac{\eta_M}{(S_{E,A}+1)^2}.
\end{align}
\begin{align}
&w_M=-\mathbf{v}_{M-1}^H\mathbf{B}_{M-1}^{-1}\mathbf{C}_{M-1}\mathbf{B}_{M|2,2}c_M\mathbf{g}_M
\notag\\
&=-t_{1,M-1}\frac{c_M^2}{S_{E,A}+1}\mathbf{g}_M^H\mathbf{B}_{M|2,2}\mathbf{g}_M\notag\\
&=-t_{1,M-1}\frac{\eta_M}{S_{E,A}+1}.
\end{align}

Therefore,
\begin{align}
&t_{1,M}=t_{1,M-1}-t_{1,M-1}^2\frac{\eta_M}{(S_{E,A}+1)^2}\notag\\
&
-2t_{1,M-1}\frac{\eta_M}{S_{E,A}+1}+\eta_M.
\end{align}

We see $t_{1,M}<t_{1,M-1}+\eta_M<\sum_{i=2}^M\eta_i$.

It follows from \eqref{eq:eta_M} that
if $S_{E,A}+1>t_{1,i-1}$ for all $i$, we have $\eta_i<c_i^2$ for all $i$ (a tight bound when $S_{E,A}$ and $S_{E,i}$ are large), and hence
\begin{equation}\label{eq:tM}
  t_{1,M}<\sum_{i=2}^M c_i^2<M-1.
\end{equation}
Here the first bound is tight when $S_{E,2}, \cdots, S_{E,M}$ and $S_{E,A}$ are large, and the second bound is tight with additional large $S_2,\cdots,S_M$.
The above suggests that if $S_{E,A}$ is sufficiently large, then $t_{1,M}<M-1$. However, as shown next, $t_{1,M}$ is an increasing function of $S_{E,A}$, and hence $t_{1,M}<M-1$ for all $S_{E,A}$.

Note that it is easy to prove $\frac{\partial t_{1,M}}{\partial S_{E,A}}
=-\mathbf{v}_M^H\mathbf{B}_M^{-1}\left (\frac{\partial}{\partial S_{E,A}}\mathbf{B}_M\right )
\mathbf{B}_M^{-1}\mathbf{v}_M>0$ where $-\frac{\partial}{\partial S_{E,A}}\mathbf{B}_M$ is positive semi-definite. It is also easy to prove $\frac{\partial }{\partial S_{E,A}}\left (\frac{t_{1,M}}{S_{E,A}+1}\right )
=-\mathbf{v}_M^H\mathbf{B}_M^{-1}\left (\frac{\partial}{\partial S_{E,A}}\left (\frac{1}{S_{E,A}+1}\mathbf{B}_M\right )\right )
\mathbf{B}_M^{-1}\mathbf{v}_M<0$ where $\frac{\partial}{\partial S_{E,A}}\left (\frac{1}{S_{E,A}+1}\mathbf{B}_M\right )$ is positive semi-definite.

It follows from \eqref{eq:q1} and \eqref{eq:tM} that
\begin{align}\label{}
  &\gamma_1-1 =c_1-q_1
  %=c_1-c_1^2\left [ \frac{t_{1,M}}{(S_{E,A}+1)^2}+\frac{S_{E,A}}{S_{E,A}+1}\right ]
  =\left (c_1-c_1^2\frac{S_{E,A}}{S_{E,A}+1}\right ) -\frac{c_1^2t_{1,M}}{(S_{E,A}+1)^2}
\end{align}
with $t_{1,M}<M-1$ for $M\geq 2$.
This along with the initial discussion after \eqref{eq:cond_1} completes the proof of Proposition \ref{proposition_gamma1}.

\subsection{Proof of proposition \ref{Proposition_symmetric}}\label{Proof_Proposition_Symmetric}
Let $\mu=\frac{\sigma^2}{1+\sigma^2}$, $\mu'=1-\mu$, $\mu_{E,A}=\frac{\sigma_{E,A}^2}{1+\sigma_{E,A}^2}$, $\mu_{E,A}'=1-\mu_{E,A}$.
It follows from \eqref{eq:AP_M} that
\begin{equation}\label{eq:inv_AP_M}
  \frac{1}{\sigma_{\Delta s_i}^2}=1+\frac{1}{g_A}
\end{equation}
with $g_A=\mu\mu'+\sigma_A^2$.

Note that $\sigma_{\Delta s_i}^2$ is invariant to $i$.
Also due to symmetry of the network,
%\begin{equation}\label{}
%  \sigma_{\Delta s_{M|E,1:M-1}}^2<\sigma_{\Delta s_{M-1|E,1:M-2}}^2
%  <\cdots <\sigma_{\Delta s_{1|E}}^2.
%\end{equation}
\begin{equation}\label{}
  \sigma_{\Delta s_{1|E}}^2>\sigma_{\Delta s_{2|E,1}}^2>\cdots>\sigma_{\Delta s_{M|E,1:M-1}}^2.
\end{equation}
So, the descending order of the terms in \eqref{eq:total_secrecy_3} is clear.

We now need to prove $\sigma_{\Delta s_{M|E,1:M-1}}^2>\sigma_{\Delta s_M}^2$ subject to a sufficient power $p_B$, we first write
\begin{align}\label{}
  &\mathbf{A}_M-\frac{1}{b}\mathbf{c}\mathbf{c}^T
  =\left [ \begin{array}{cc}
              \mathbf{B}_M & \mathbf{d} \\
              \mathbf{d}^T & a_E
            \end{array}
  \right ]
\end{align}
where $a_E=1+\mu'+\sigma_E^2-\mu_{E,A}'^2\mu'^2$, $\mathbf{d}=\mu_{E,A}\mu'^2\mathbf{1}$ and
\begin{equation}\label{}
  \mathbf{B}_M = \mu_{E,A}\mu'^2\mathbf{1}\mathbf{1}^T+(a_E-1-\mu_{E,A}\mu'^2)\mathbf{I}.
\end{equation}
Applying $(\mathbf{I}+\mathbf{x}\mathbf{x}^H)^{-1}=\mathbf{I}-\frac{1}{1+\|\mathbf{x}\|^2}\mathbf{x}
\mathbf{x}^H$ to \eqref{eq:EEE} with $i=M$, one can verify that
\begin{align}\label{eq:inv_EEE}
&\frac{1}{\sigma_{\Delta s_{M|E,1:M-1}}^2}
=1+\frac{1}{g_E}
\end{align}
with
\begin{equation}\label{}
  g_E=\sigma_E^2+\mu'-\mu'_{E,A}\mu'^2-
  \frac{(M-1)\mu_{E,A}^2\mu'^3}{\frac{a_E^2}{\mu'}+\mu+(M-1)\mu_{E,A}\mu'}
\end{equation}

To compare \eqref{eq:inv_AP_M} and \eqref{eq:inv_EEE}, we consider
\begin{align}
&g_E-g_A=\sigma_E^2-\sigma_A^2+
\mu'\frac{\mu_{E,A}\sigma_E^2+\mu_{E,A}\mu'\mu}
{\frac{\sigma_E^2}{\mu'}+\mu+(M-1)\mu_{E,A}\mu'}.
\end{align}

It is obvious that if $\sigma_E^2\geq \sigma_A^2$, we have $g_E-g_A>0$ and hence $R_{s,M}'>0$ for any (positive) power $p_B$ used by each UE.

Now consider the case of $\sigma_E^2< \sigma_A^2$ or $\beta_0 = \frac{\sigma_A^2}{\sigma_E^2}>1$. Then $g_E-g_A>0$ if and only if
\begin{equation}\label{}
  c_2\sigma_A^4+c_1\sigma_A^2-c_0<0
\end{equation}
or equivalently,
\begin{equation}\label{}
  \sigma_A^2<\frac{1}{2}\left (-\frac{c_1}{c_2}+\sqrt{\left (\frac{c_1}{c_2}\right )^2
  +4\frac{c_0}{c_2}}\right )\doteq \bar\sigma_A^2
\end{equation}
where $c_2=\frac{\beta_0-1}{\beta_0^2\mu'^2}$, $c_1=\frac{(\beta_0-1)(\mu+(M-1)\mu_{E,A}\mu')}{\beta_0\mu'}-\frac{\mu_{E,A}}{\beta_0}$, and $c_0=\mu_{E,A}\mu\mu'$.
Here $\bar\sigma_A^2$ is positive and invariant to $\sigma_A^2$ while $\sigma_A^2$ is inversely proportional to the power $p_B$ used by each UE in phase 2. So, we have proven that subject to any given $\beta_0>1$, there is a power threshold $\bar p_B$ such that the smallest term $R_{s,M}'$ in \eqref{eq:total_secrecy_3} is positive when $p_B>\bar p_B$.

For $\beta_0>1$ and a large $M$, we have $c_1=\mathcal{O}(M)$ and hence
\begin{align}
&\bar\sigma_A^2\approx
\frac{1}{2}\left (-\frac{c_1}{c_2}+\frac{c_1}{c_2}\left (1+\frac{1}{2}\frac{4c_0/c_2}{c_1^2/c_2^2}\right )\right )
=\frac{c_0}{c_1}\notag\\
&=\mathcal{O}\left (1/M\right ).
\end{align}
In this case, $\bar p_B$ increases linearly with $M$.

\end{appendices}

% \section*{Acknowledgement}
%The author would like to express his appreciation of the comments by anonymous reviewers of \cite{Hua2023Sept}. Special thanks also goes to Dr. P. Liang of RF-DSP, Dr. W. Yang of Qualcomm and Dr. J. Perazzone of ARL for their comments and discussions on STEEP.

\end{document}